\documentstyle[preprint,aps]{revtex}
\newcommand{\vp}{{\bf p}}
\newcommand{\al}{\alpha}
\newcommand{\bt}{\beta}

\newcommand{\be}{\begin{equation}}
\newcommand{\ee}{\end{equation}}
\newcommand{\ba}{\begin{eqnarray}}
\newcommand{\ea}{\end{eqnarray}}
\newcommand{\r}{{\bf r}}
\newcommand{\up}{\stackrel{<}{>}}
\newcommand{\down}{\stackrel{>}{<}}
\newcommand{\co}{({\bf R},T)}

\newcommand{\cof}{({\bf p},{\bf R},T)}
\newcommand{\nee}{\tilde{\epsilon}_{p}}
\def\pomrt{(\vp,\omega;{\bf R},T)}

\begin{document}

\title{Quasiparticle kinetic equation in a 
trapped Bose gas at low temperatures}

\author{M. Imamovic-Tomasovic and A. Griffin}

\address{Department of Physics, University of Toronto \\ 
Toronto, Ontario, Canada M5S 1A7 }

\maketitle
\begin{abstract}
Recently the authors used the Kadanoff-Baym non-equilibrium 
Green's function formalism to  
 derive kinetic equation for the non-condensate atoms, in conjunction 
 with a consistent generalization of the Gross-Pitaevskii equation 
 for the Bose condensate wavefunction. This work was 
limited to 
 high temperatures, where the excited atoms could be described by a 
 Hartree-Fock particle-like spectrum. We present the 
generalization of this recent work to low temperatures, where the 
single-particle spectrum is now described by  
the Bogoliubov-Popov approximation. We derive a kinetic equation for 
the quasiparticle distribution function with collision integrals 
describing scattering between quasiparticles and the condensate atoms.
From the general expression for the collision integral for the 
scattering between quasiparticle excitations, we find 
the quasiparticle distribution function corresponding to 
local equilibrium. This 
expression includes a quasiparticle chemical potential that 
controls the non-diffusive equilibrium between condensate atoms and 
the 
quasiparticle excitations. 
We also derive a generalized Gross-Pitaevskii equation for the 
condensate wavefunction that includes the damping effects due to 
collisions between atoms in the condensate and 
the thermally excited  quasiparticles.
 For a uniform Bose gas, 
our kinetic equation for the thermally excited quasiparticles 
reduces to that found by Eckern as well as Kirkpatrick and Dorfman.
\end{abstract}

\section{Introduction}
In a trapped, weakly-interacting Bose gas at 
{\it T}=0, the fraction of atoms that are excited out 
of the condensate is only a few percent \cite{strtheory}. 
As a result, the dynamics 
of the trapped Bose gas at low temperatures 
(compared to $T_{BEC}$) is well described 
by the equation of motion for the macroscopic 
wavefunction $\Phi({\r},t)$. This is the 
time-dependent  Gross-Pitaevskii equation
(GP) \cite{strtheory}
\be
i\frac{\partial \Phi({\r},t)}{\partial t}
=\left[-\frac{1}{2m}\nabla_{\r}^{2}
+U_{ext}({\r})+gn_{c}({\r},t)\right]\Phi({\r},t),
\label{eq:gpeq}
\ee
where $n_{c}({\r},t)=|\Phi({\r},t)|^{2}$ is the 
non-equilibrium density of the atoms in the condensate 
and $U_{ext}({\r})$ is a harmonic trap potential 
(In this paper, we set $\hbar=1$). 
For a discussion of the properties of a dilute Bose gas at very low 
temperatures, only the {\it s}-wave component of the two-body 
interaction 
$v({\bf r}-{\bf r'})$ is important. Thus one can use the 
pseudopotential  
\be
v({\bf r})=g\delta({\bf r}), \hspace{5mm} g=4\pi a/m, 
\label{eq:continteraction}
\ee
where {\it a} is the {\it s}-wave scattering length of the true 
potential. The GP equation describes the 
motion of the condensate moving in the dynamic Hartree 
mean-field produced by the other atoms in the condensate
and gives a closed equation for the order parameter 
$\Phi({\r},t)$. 
The GP equation (\ref{eq:gpeq})  provides a very accurate 
description of the static
and dynamic properties of a trapped Bose gas at low 
temperatures $T\leq 0.4 \hspace{1mm}T_{BEC}$, as confirmed by many 
experiments in the last few years \cite{strtheory}. 
In superfluid ${}^{4}$He, the non-condensate 
fraction at {\it T}=0 is close to 90\% \cite{sokolBEC}.  
Thus in superfluid ${}^{4}$He, one always has to deal with {\it both} 
the condensate and non-condensate atoms. Clearly a closed GP equation 
for $\Phi({\r},t)$ like (\ref{eq:gpeq}) is never valid in 
superfluid ${}^{4}$He. 

At finite temperatures (say $T>0.5 \hspace{1mm}T_{BEC}$), however, 
the number 
of atoms 
thermally  excited out of the condensate becomes 
significant and the GP equation (\ref{eq:gpeq})
 is no longer sufficient. 
The simplest way to include the effect of the excited 
atoms on the condensate is to add the additional 
Hartree-Fock mean field  $V_{HF}=2g\tilde{n}({\r},t)$ 
produced by the non-condensate atoms (here $\tilde{n}$ 
is the local non-condensate density). One 
immediately sees this new GP equation is no longer 
closed since it depends on the dynamics of 
the non-condensate atoms. 

To find the time-dependent non-condensate density, 
Zaremba, Nikuni and Griffin (ZNG) \cite{zngjltp}
have used a {\it quantum Boltzmann equation} for the 
single-particle distribution function of the non-condensate atoms 
$f({\bf p},\r,t)$
\be
\left[\frac{\partial}{\partial t}+\frac{\bf p}{m}
\cdot \nabla_{\r}-\nabla_{\r}U({\r},t)\cdot 
\nabla_{\bf p}\right]f({\bf p},\r,t)
=\left[\frac{\partial f({\bf p},\r,t)}
{\partial t}\right]_{coll}. \label{eq:kinht}
\ee 
Here, the thermally excited atoms are assumed to be well 
described by the single-particle spectrum 
$\frac{p^{2}}{2m}+U({\r},t)$, where
\be
U({\r},t)\equiv U_{ext}(\r)+2g\left[n_{c}({\r},t)
+\tilde{n}({\r},t)\right]
\label{eq:effectivefield}
\ee
includes the self-consistent Hartree-Fock dynamic mean field
involving the {\it total} time-dependent local density 
$n({\r},t)$. The right-hand side of (\ref {eq:kinht})  
describes the effect of collisions between atoms on 
the time evolution of the distribution function 
$f({\bf p},\r,t)$. In Bose-condensed gases, 
this collision integral has two distinct contributions
\be
\left[\frac{\partial f}{\partial t}\right]_{coll}
=C_{12}[f]+C_{22}[f].
\label{eq:collint}
\ee
Here, $C_{22}$ denotes the part of the collision 
integral that describes two-body collisions between 
non-condensate atoms. Above $T_{BEC}$, this is the 
only term present. In contrast, $C_{12}$ describes collisions 
involving non-condensate atoms and  
{\it one} condensate atom. The role of $C_{12}$ is 
crucial since it couples the condensate and 
non-condensate components. 

ZNG also derived
a generalized Gross-Pitaevskii equation that includes the effect of 
the collisions between the atoms in the condensate and the thermal 
cloud. Recently, we obtained the equations of motion derived 
by ZNG at finite temperatures in a more elegant way using the 
well-known Kadanoff-Baym Green's functions formalism \cite{itg2}. 
We also note that several other groups have also recently 
discussed the finite temperature 
dynamics of a trapped Bose-condensed gas 
\cite{stoofjltp,walser,gardiner,nicknew}, 
each using a somewhat different formalism.

The kinetic equation (\ref{eq:kinht}) is valid 
in the semiclassical limit only: it assumes that 
the thermal energy is much greater than the spacing 
between the trap SHO energy levels ($k_{B}T\gg \hbar 
\omega_{0}$, where $\omega_{0}$ is the harmonic 
well frequency) as well as the average interaction energy 
($k_{B}T\gg gn$). ZNG have given a detailed 
derivation \cite{zngjltp} of (\ref {eq:kinht}) 
at finite temperatures for a trapped Bose gas 
using the approach of Kirkpatrick 
and Dorfman \cite{kd}, who considered 
a uniform Bose gas. However, the ZNG theory is not applicable to very 
low temperatures because the thermal excitations 
on which it is based do 
not include the collective (or phonon) part 
of the Bogoliubov spectrum 
$E_{p}=\sqrt{(p^{2}/2m)^{2}+gn_{c}p^{2}/m}$. 
To make this generalization, one has to formulate a kinetic 
theory in terms of such Bogoliubov quasiparticle excitations. 
In the paper, we use the Kadanoff-Baym (KB) 
non-equilibrium Green's functions method \cite{kb} to 
derive such a generalized kinetic equation for the thermally excited 
Bogoliubov quasiparticles. To do this, we work within the 
second-order  
Beliaev-Popov  approximation \cite{shi}. 

Kadanoff and Baym first formulated the method of 
deriving a kinetic equation for a normal interacting system 
using non-equilibrium Green's functions \cite{kb}.
Kane and Kadanoff (KK) \cite{kk,kt} generalized this method 
to deal with a Bose-condensed gas, with the specific goal of using 
the 
resulting kinetic equations to derive the two-fluid 
hydrodynamics equations of Landau \cite{khalatnikov}. 
An excellent review of the nonequilibrium real-time 
Green's functions and the generalized 
kinetic equation for the normal systems (non Bose-condensed) 
can be found in the book by Zubarev, Morazov and Ropke 
\cite{zubarevbook}. 

It is important to 
emphasize that although our analysis involves the non-equilibrium 
generalization of the Beliaev second-order self-energy  
used for systems in thermal equilibrium, our work is 
quite different from the recent papers discussing the poles of 
equilibrium Green's functions within the 
second-order Beliaev approximation \cite{shi,giorgini99,fedichev}. 
In a very elegant formulation, Giorgini
\cite{giorgini99} has calculated the quasiparticle energy and damping 
at finite temperatures in a dilute Bose gas in the 
collisionless regime by linearizing the equations of motion for 
fluctuations using the first-order dynamic 
Hartree-Fock-Bogoliubov approximation. This leads,
as expected,  to the same excitations spectrum 
found by  Shi and Griffin \cite{shi}
who calculated directly the poles of the single-particle 
equilibrium Green's functions using the second-order Beliaev 
self-energy diagram contributions.  

In the present paper, we use the second-order Beliaev approximation 
to discuss the {\it non-equilibrium dynamics} of a 
trapped Bose-condensed gas at finite temperatures. We use 
the second-order Beliaev self-energies with the lower order 
Bogoliubov 
excitation spectrum, including off-diagonal single-particle 
propagators, 
but we ignore the anomalous correlation function $\tilde{m}$.
This last assumption defines what we call the 
Bogoliubov-Popov approximation \cite{hfb}. 
It is important to note that, in this 
paper, we are primarily interested in the damping effect arising 
from the collisions between atoms. We do  not calculate  the 
second-order corrections in $g$ to the quasiparticle spectrum or to 
the condensate chemical potential that 
are associated with the real parts of the second-order Beliaev 
self-energies. One can show that the real part of the second-order 
self-energies enter into the kinetic equation mainly through 
renormalized 
quasiparticle energy (see Eq. (6.3.77) in Ref. 
\cite{zubarevbook}).  This suggests a simple way of extending the 
kinetic equations we derive by using an improved quasiparticle 
spectrum.

We derive the 
kinetic equation for the distribution function for the 
thermally excited quasiparticles, as well as a 
generalized equation for the Bose condensate order parameter. 
The kinetic equation we obtain is the same as the one 
derived for a uniform Bose gas by Eckern \cite{eckern} in 1984, 
and by by Kirkpatrick and 
Dorfman (KD) in 1985 \cite{kd}. The KD 
derivation was based on a direct extension of the traditional method
 used to derive kinetic equations for 
classical gases, which obscured much of the physics. Moreover,  KD 
did not explicitly
derive equations of motion for the condensate degree of freedom.
In a non-Bose condensed uniform gas, similar kinetic equations are 
derived in Ref. \cite{zubarevbook} using the related Keldysh 
formalism.

The kinetic equation for the quasiparticle excitations which  we 
derive in this paper, coupled to a generalized GP equation, 
provides a platform for studying different non-equilibrium aspects of 
a 
dynamics of a trapped, Bose-condensed gas at all temperatures, both 
in the collisionless and hydrodynamic domains. 
Linearizing our equations of motion around static equilibrium, 
one could calculate the density 
response functions which would exhibit collective mode resonances 
with a spectrum which goes past the generating Beliaev approximation. 
The high temperature limit of these coupled equations 
\cite{zngjltp,itg2} has been recently used 
to study the dynamics of the condensate formation and growth in an 
inhomogeneous, trapped Bose gas \cite{BZStoof}.
By taking the moments of the kinetic 
equation, the  two-fluid 
hydrodynamic equations have been derived \cite{zngjltp,nikuninew}. 

The present paper is a natural generalization of our two earlier 
papers 
based on the KB formalism. In Ref. \cite{itg1}, we have derived 
kinetic equations within the full Hartree-Fock-Bogoliubov 
approximation but ignored 
collisions. In Ref. \cite{itg2}, we derived a kinetic equation 
including collisions, but which was based on a simple Hartree-Fock 
particle-like spectrum and hence was not valid at very low 
temperatures.

In Section II, we review the general equations of motion for the 
non-equilibrium Green's functions describing the non-condensate atoms 
as well as the equation of motion for the macroscopic order 
parameter. In 
Section III, we transform these equations to a local rest 
frame of reference 
where the order parameter $\Phi(\r,t)$ is real, i.e., to a frame 
where 
the local superfluid velocity is zero. 
This naturally introduces the superfluid velocity 
and the local chemical potential as the spatial and time 
derivatives, respectively,  of the phase of the order parameter 
\cite{kk}.  We then specialize our equations of motion for the 
non-equilibrium Green's functions for the case 
of slowly varying external perturbations. We use the key 
assumption that all correlation functions vary 
slowly as a function of center-of-mass 
space-time coordinates but are dominated by small values of the 
relative 
coordinates. Following the Kadanoff-Baym approach \cite{kb,kk,kt}, 
we derive a generalized quantum Boltzmann equation for 
the frequency-dependent quasiparticle distribution 
function $f\pomrt$. 

In Section IV, we 
use this generalized KB quantum Boltzmann equation to derive the 
kinetic equation for quasiparticles at low temperatures, 
with the collision integral describing collisions between 
quasiparticles in the Bogoliubov-Popov 
approximation. In Section V, we discuss a general form of the 
local quasiparticle distribution function and introduce a new 
quasiparticle 
chemical potential situations in which  
the condensate and thermal excitations are not in diffusive 
equilibrium. 
In Section VI, we derive, in a self-consistent manner, a
generalized 
Gross-Pitaevskii equation. In Section VII, we verify that our 
coupled equations exhibit the Kohn mode corresponding to the harmonic 
oscillations of the center-of-mass of the equilibrium condensate and 
non-condensate density profiles.

\section{Equations of motion for non-equilibrium 
 Green's functions}

For convenience, we first review the KB formalism \cite{kb} 
already used in our earlier work \cite{itg2,itg1}. 
In terms of quantum field operators, the many-body 
Hamiltonian ($\hat{K}=\hat{H}-\mu_{0}\hat{N}$) 
describing interacting Bosons confined 
by an external harmonic potential 
$U_{ext}({\bf r})$ is given by:
\ba
\hat{K}&=&\int d{\bf r} {\psi}^{\dag}(\r)
\left[-\frac{1}{2m}\nabla^{2}_{\r}+U_{ext}(\r)
-\mu_{0}\right]{\psi}(\r) \nonumber \\
&+&\frac{1}{2} \int d\r d\r'
{\psi}^{\dag}(\r){\psi}^{\dag}(\r') 
v(\r-\r'){\psi}(\r){\psi}(\r').
\ea
We separate out the condensate part of the field 
operator in the usual fashion\cite{fw,beliaev58} 
\begin{equation}
\psi(\r)=\langle\psi(\r)\rangle_{t}+\tilde{\psi}(\r),
\label{eq:seppsi}
\end{equation}
where $\langle \tilde{\psi}(\r)\rangle=0$ and 
$\langle\psi(\r)\rangle_{t}=\Phi(\r,t) $ is the 
Bose macroscopic wavefunction. The non-condensate 
(or excited-atom component) field operators  
$\tilde{\psi}(\r)$ and $\tilde{\psi}^{\dag}(\r)$ 
satisfy the usual Bose commutation relations.

In a Bose-condensed 
system, the finite value of $\Phi(\r,t)$ leads 
to finite values of the off-diagonal 
(or anomalous) propagators 
$\langle\tilde{\psi}(1)\tilde{\psi}(1')\rangle$ 
and 
$\langle\tilde{\psi}^{\dag}(1)
\tilde{\psi}^{\dag}(1')\rangle$. These must be dealt
with on an equal basis with the diagonal (or normal) 
propagators, and thus it is convenient to work with a single-particle 
$2\times 2$ matrix Green's function 
defined by \cite{kk,hm}
\begin{equation}
\hat{g}(1,1';U)=-i\left( \begin{array}{cc}\langle T 
\psi(1) \psi^{\dag}(1')\rangle\hspace*{5mm}\langle T 
\psi(1) \psi(1')\rangle\\ 
\langle T \psi^{\dag}(1) \psi^{\dag}(1')\rangle
\hspace*{5mm} \langle T\psi^{\dag}(1)\psi(1')\rangle 
\end{array}\right). \label{eq:deftildeg}
\end{equation}
Here, {\it T} represents the time-ordering 
operator and we use the usual KB abbreviated notation, 
$1 \equiv (\r,t)$ and $1'\equiv (\r',t'$). We 
define $\hat{g}^{<}$ and $\hat{g}^{>}$ by
\ba
\hat{g}(1,1';U)&=&\hat{g}^{>}(1,1';U)\hspace{10mm}
t_{1}>t_{1'} \nonumber \\
&=&\hat{g}^{<}(1,1';U)\hspace{10mm}   t_{1}<t_{1'}.
\label{eq:glg}
\ea
Using (\ref{eq:seppsi}), the matrix propagator in 
(\ref{eq:deftildeg})  splits into two parts
\be
\hat{g}(1,1';U)=\hat{\tilde{g}}(1,1';U)+\hat{h}(1,1';U).
\ee
Here $\hat{\tilde{g}}$ is identical to 
(\ref{eq:deftildeg}), except that it involves the 
non-condensate part of the field operators. 
The condensate propagator is given by
\be
\hat{h}(1,1';U)\equiv -i\left ( \begin{array}{cc} 
\Phi(1)\Phi^{*}(1')   \hspace*{5mm} \Phi(1) 
\Phi(1') \\ \Phi^{*}(1)\Phi^{*}(1')
\hspace*{5mm} \Phi^{*}(1)\Phi(1') 
\end{array} \right ),
\label{eq:defh}
\ee
with $\langle\psi^{\dag}(\r)\rangle_{t}
\equiv \Phi^{*}(\r,t)$. 

A very useful and elegant way of generating the equations
of motion for both $\hat{\tilde{g}}$
and $\Phi$ is to use functional derivatives with 
respect to weak external fields \cite{kb,hm,cheung}, 
\be
H'(t_{1})=\frac{1}{2}\int d{\bf r}_{1} 
d2{\psi}^{\dag}(1)U(1,2){\psi}(2) 
+\int d {\bf r}_{1}\left[\psi^{\dag}(1)\eta_{ext}(1)
+\psi(1)\eta^{*}_{ext}(1)\right].
\label{eq:externalh}
\ee
Here $U(1,2) $ is an external generating scalar 
field non-local in space and time. 
It represents a perturbation in which a particle 
is removed from the system at point 1 and 
added at 2. The symmetry-breaking fields 
$\eta_{ext}$ and $\eta^{*}_{ext}$ describe 
particle creation and destruction \cite{hm,bog}. 
All higher-order Green's functions can be neatly expressed 
as functional derivatives of single-particle 
Green's functions with respect to such generating fields.

Following the  Kane-Kadanoff (KK) analysis  \cite{kk,kb}, 
the Dyson-Beliaev equations of motion  for the real-time 
non-condensate propagators  $\hat{\tilde{g}}(1,1')$ 
can be written in the following 
$2 \times 2$ matrix form
\ba 
&&\int d\bar{1}\left[{\hat{g}_{0}}^{-1}(1,\bar{1})-
\hat{\Sigma}^{HF}(1, \bar{1})\right] 
\hat{\tilde{g}}^{\up}(\bar{1},1')\nonumber \\
 &=&\int_{-\infty}^{t_{1}}d\bar{1}\hat{\Gamma}(1,\bar{1})
\hat{\tilde{g}}^{\up}(\bar{1},1')
-\int_{-\infty}^{t_{1'}} d\bar{1}\hat{\Sigma}_{c}^{\up}
(1,\bar{1})\hat{a}(\bar{1},1'),
\label{eq:tildeg2}
\ea 
and 
\ba 
&&\int d\bar{1}\hat{\tilde{g}}^{\up}(1,\bar{1})
\left[{\hat{g}_{0}}^{-1}(\bar{1},1')-
\hat{\Sigma}^{HF}( \bar{1},1')\right] \nonumber \\
&=&\int_{-\infty}^{t_{1}}d\bar{1}\hat{a}(1,\bar{1})
\hat{\Sigma}_{c}^{\up}(\bar{1},1')
-\int_{-\infty}^{t_{1'}}d\bar{1}\hat{\tilde{g}}^{\up}
(1,\bar{1})\hat{\Gamma}(\bar{1},1').
\label{eq:tildeg3}
\ea 
Here $\hat{a}(1,1')$ and $\hat{\Gamma}(1,1')$ are 
defined by the matrix elements
\ba
&&a_{\al \bt}(1,1')\equiv 
\tilde{g}_{\al \bt}^{>}(1,1')
-\tilde{g}_{\al \bt}^{<}(1,1') \nonumber \\
&&\Gamma_{\al \bt}(1,1')\equiv 
\Sigma_{\al \bt}^{>}(1,1')
-\Sigma_{\al \bt}^{<}(1,1').
\label{eq:defagamma}
\ea
The {\it non-equilibrium} single-particle spectral density 
$a_{\al \bt}(1,1')$ will play 
a crucial role in this paper. In the above 
equations and elsewhere, integration over $d{\bar 1}$ 
means integration  over the coordinates 
$({\bf r}_{1}, t_{1})$ and a trace over the 
matrix index $\alpha_{1}$; and
$\delta(11')\equiv\delta({\r}-{\r'})\delta(t-t')$.

The single-particle self-energy which 
is involved in (\ref{eq:tildeg2}) and (\ref{eq:tildeg3})
has already been split into two parts \cite{kk,hm}
\be
\hat{\Sigma}(1,1')=\hat{\Sigma}^{HF}(1,1')
+\hat{\Sigma}_{c}(1,1').
\ee
The first-order Hartree-Fock self-energies are given by
\be
\hat{\Sigma}^{HF}(11')=g\left ( \begin{array}{cc} 
2n(1),  \hspace*{3mm} m(1) \\ 
m^{*}(1),\hspace*{3mm} 2n(1) \end{array} 
\right )\delta(11'), \label{eq:sigmahf}
\ee
and $\Sigma_{c}$ is the  second-order ``collisional''  
Beliaev self-energy. The total density is given by 
$n(1)\equiv i\tilde{g}^{<}_{11}(1,1^{+})+|\Phi(1)|^{2}
=\tilde{n}(1)+n_{c}(1)$ and the total anomalous density by 
$m(1)\equiv i\tilde{g}_{12}(1,1)+[\Phi(1)]^{2}$. 
In addition, we define (see (\ref{eq:glg}))
\ba
\hat{\Sigma}_{c}(1,1')&=&\hat{\Sigma}_{c}^{>}(1,1')
\hspace{10mm} t_{1}>t_{1'} \nonumber \\
&=&\hat{\Sigma}_{c}^{<}(1,1')\hspace{10mm}t_{1}<t_{1'}.
\ea 
In (\ref{eq:tildeg2}) and (\ref{eq:tildeg3}), 
the inverse of the non-interacting $2\times 2$ matrix 
Bose gas propagator $\hat{g}_{0}(1,1')$ is defined by
\be
\hat{g}_{0}^{-1}(1,1')=\left[i{\bf \tau}_{3}
\frac{\partial}{\partial t_{1}}
+\frac{\nabla_{1}^{2}}{2m}-U_{ext}(\r_{1})+\mu_{0}
\right] \delta(1,1').
\label{eq:godef}
\ee
 
We note that the equations in (\ref{eq:tildeg2}) 
and (\ref{eq:tildeg3}) already have the ``structure'' 
of a kinetic equation such as (\ref{eq:kinht}). 
The Hartree-Fock part of the self-energy has been
included into the left-hand side of (\ref{eq:tildeg2}) 
and (\ref{eq:tildeg3}), giving the mean-field 
contribution to the ``streaming'' term. 
The second-order self-energy describing binary collisions is 
included on the right-hand side of (\ref{eq:tildeg2}) and 
(\ref{eq:tildeg3}), and it will be shown to 
give rise to collision integrals in the quasiparticle 
 kinetic equation we derive.  

In this paper, 
we work with the second-order self-energy $\Sigma_{c}$ 
as given by 
the Beliaev (gapless) approximation \cite{shi,hfb,hm}. 
The advantage of the Beliaev approximation is that 
the non-condensate single-particle Green's function exhibits the 
correct quasiparticle spectrum (phonon-like in the 
long-wavelength, uniform gas limit). 
In the second-order Beliaev approximation, the second-order 
self-energy 
$\Sigma^{\down}$  is given by \cite{itg2}
\ba
\hat{\Sigma}_{c}^{\down}(1,1')=&-&\frac{1}{2}v(13)v(21')
\tilde{g}^{\down}(11')\left[\tilde{g}^{\up}(23)
\tilde{g}^{\down}(32)+\tilde{g}^{\up}(23)h(32)
+h(23)\tilde{g}^{\down}(32)\right]
\nonumber \\
&-&v(13)v(21')\tilde{g}^{\down}(12)\left[
\tilde{g}^{\up}(23)h(31')+
h(23)\tilde{g}^{\down}(31')+\tilde{g}^{\up}(23)
\tilde{g}^{\down}(31')\right] \nonumber \\
&-&\frac{1}{2}v(13)v(21')\left[h(11')\tilde{g}^{\up}(23)
\tilde{g}^{\down}(32)+2h(12)\tilde{g}^{\up}(23)
\tilde{g}^{\down}(31')\right].
\label{eq:sigmabeliaev}
\ea

The equation for the condensate can be 
written (see Ref. \cite{itg2} for more details) in terms of 
the 2-component order parameter 
$\hat{G}_{1/2}(1)\equiv\sqrt{-i}\langle\Psi(1)\rangle$, 
where $\Psi$ is defined as
\be
\Psi(1)\equiv \left(\begin{array}{c} \psi(1) \\  
\psi^{*}(1) \end{array}\right). 
\ee
One finds that equation of motion for the $\hat{G}_{1/2}$ 
is given by \cite{itg2,hm,cheung} 
\be
\int d\bar{1}{\hat{g}_{0}}^{-1}(1,\bar{1})
G_{1/2}(\bar{1})=\sqrt{-i}\hat{\eta}(1)
+\sqrt{-i}\hat{\eta}_{ext}(1),
\label{eq:orderg1/21}
\ee 
where the so-called condensate source function 
$\eta$ is defined by the three-field 
correlation function 
\be
\sqrt{-i}\hat{\eta}(1) \equiv \frac{1}{2}
\int d\bar{2} \sqrt{-i}v(1\bar{2})\langle
T\Psi(1)\Psi^{\dag}(\bar{2})\Psi(\bar{2})\rangle.
\label{eq:etadef}
\ee
The external particle-source fields are defined 
in (\ref{eq:externalh}), with
\be
\hat{\eta}_{ext}(1) \equiv \left( \begin{array}{c} 
\eta_{ext}(1) \\ 
\eta^{*}_{ext}(1) \end{array}\right).
\ee
The exact coupled equations of motion (\ref{eq:tildeg2}),
(\ref{eq:tildeg3}) and 
(\ref{eq:orderg1/21}) are the starting point of our analysis.
The external generating fields $U$ and $\eta_{ext}$ 
will be left implicit in the rest of this paper.

Using (\ref{eq:seppsi}), one can also decompose
the three-field correlation function involved in $\eta$ 
defined in (\ref{eq:etadef}). For example, one has
\ba
\langle T \psi(1)\psi^{\dag}(2)\psi(2)\rangle&=&
\Phi(1)n_{c}(2)+\Phi(2)\langle T\tilde{\psi}(1)
\tilde{\psi}^{\dag}(2)\rangle \nonumber \\ 
&+&\Phi^{*}(2)\langle T\tilde{\psi}(1)\tilde{\psi}(2)\rangle
+\langle T\tilde{\psi}(1)
\tilde{\psi}^{\dag}(2)\tilde{\psi}(2)\rangle.
\label{eq:threefield}
\ea
In the first order Hartree-Fock-Bogoliubov (HFB) approximation,
one neglects the three-field correlation function 
$\langle \tilde{\psi} \tilde{\psi}^{\dag}\tilde{\psi}\rangle$
for the non-condensate atoms. 
In this approximation, (\ref{eq:threefield}) only involves 
the condensate density $n_{c}$ and the two-field correlation 
functions $\tilde{n}$ and $\tilde{m}$, as defined earlier. In this 
paper, in contrast, we keep the non-condensate three-field 
correlations but  
will eventually set $\tilde{m}\equiv\langle \tilde{\psi}(1)
\tilde{\psi}(1)\rangle$ (the Popov approximation \cite{hm,hfb}).
    
Equation (\ref{eq:orderg1/21}) can be rewritten in terms of a 
condensate self-energy function $S$, defined by
\be
\int d\bar{1}S(1,\bar{1})h(\bar{1},1')\equiv\sqrt{-i}
\eta(1)G_{1/2}^{\dag}(1'), \label{eq:defcse}
\ee
with the condensate propagator $h$ given in (\ref{eq:defh}).
In place of (\ref{eq:orderg1/21}), we have  
\ba
\int d\bar{1}\left[{\hat{g}_{0}}^{-1}(1,\bar{1})
-S^{HF}(1,\bar{1})\right]
h(\bar{1},1')&=&\int_{-\infty}^{t}d\bar{1}
(S^{>}(1,\bar{1})-S^{<}(1,\bar{1}))h(\bar{1},1')
\label{eq:orderg1/2}
\ea 
where, as before, the mean-field contributions 
(see (\ref{eq:threefield})) 
are included in the Hartree-Fock part of the 
condensate self-energy $S^{HF}$ 
\be
\hat{S}^{HF}(11')=g\left ( \begin{array}{cc} 
n(1)+\tilde{n}(1),  \hspace*{3mm} \tilde{m}(1) \\ 
\tilde{m}^{*}(1),\hspace*{3mm} n(1)+\tilde{n}(1) \end{array} 
\right )\delta(11'). \label{eq:shf}
\ee
The functions $S^{\down}$ on the right-hand side of 
(\ref{eq:orderg1/2}) contain the
second-order contributions to the 
condensate self-energy \cite{itg2}
\ba
\hat{S}^{\down}(1,1')=&-&\frac{1}{2}v(13)v(21')
\tilde{g}^{\down}(11')\left[\tilde{g}^{\up}(23)
\tilde{g}^{\down}(32)\right]\nonumber \\
&-&v(13)v(21')\tilde{g}^{\down}(12)\left
[\tilde{g}^{\up}(23)\tilde{g}^{\down}(31')\right]. 
\label{eq:sigmacondensate}
\ea
As we will show in Section VI, these contributions give rise to  
a  dissipative term in a generalized GP equation. 
If we recall the definition for the condensate propagator 
$h$ in (\ref{eq:defh}) and use the explicit form for the condensate 
Hartree-Fock self energy $S^{HF}$ in (\ref{eq:shf}), we obtain 
a generalized GP equation for the macroscopic wave function 
$\Phi(\r,t)$, namely
\ba
&&\left[-\frac{\partial}{\partial t_{1}}+\frac{\nabla_{1}^{2}}
{2m}-U_{ext}(\r_{1})+\mu_{0}-g\left(n_{c}(1)+2\tilde{n}(1)
\right)\right]\Phi(1)=g\tilde{m}(1)\Phi^{*}(1)
\nonumber \\  
&+&\int_{-\infty}^{t}d{\bar 1}\left(S_{11}^{>}
-S_{11}^{<}\right)(1,\bar{1})\Phi(\bar{1})+
\int_{-\infty}^{t}d{\bar 1}\left(S_{12}^{>}
-S_{12}^{<}\right)(1,\bar{1})\Phi^{*}(\bar{1}).
\label{eq:explicitgp}
\ea
Initially, the equation of motion for the macroscopic order 
parameter (\ref{eq:orderg1/21}) was given in terms of the condensate 
three-field source function $\eta$ in (\ref{eq:etadef}). 
We have rewritten the equation for $\Phi$ in 
terms of the condensate self-energies $S$ defined by (\ref 
{eq:defcse}) 
because, as we will see in Section VI, $S$ is more convenient 
to work with than the condensate source 
function $\eta$. 

\section{Generalized kinetic equation}

In general, the order parameter $\Phi(\r,t)$ in a Bose fluid 
is complex. It is often written 
in terms of the condensate amplitude and phase
\be
\Phi(\r,t)=\sqrt{n_{c}(\r,t)}e^{i\theta(\r,t)}.
\label{eq:31}
\ee
To derive a generalized kinetic equation, the rapid oscillations of 
the phase in (\ref{eq:31}) cause difficulties.
Following KK,  we first gauge transform 
(\ref{eq:tildeg2}) and (\ref{eq:tildeg3}) 
to the local rest frame in which the superfluid velocity is zero. 
This corresponds to removing the phase of the 
macroscopic wavefunction.
 The required gauge transformations on 
$\hat{h}(1,1')$ and $\hat{\tilde{g}}^{\up}¥(1,1')$
are \cite{kk}
\ba
\hat{h'}(1,1')&=&e^{-i\theta(1){\bf 
\tau^{(3)}}}\hat{h}(1,1')e^{i\theta(1){\bf \tau^{(3)}}}, \nonumber \\
\hat{\tilde{g'}}^{\up}¥(1,1')&=&e^{-i\theta(1){\bf 
\tau^{(3)}}}\hat{\tilde{g}}^{\up}¥(1,1')
e^{i\theta(1){\bf \tau^{(3)}}}, 
\label{eq:gauge}
\ea
where ${\bf \tau^{(3)}}$ is the Pauli spin matrix.
The physical interpretation of (\ref{eq:gauge}) is that 
it involves a transformation to a coordinate system in which 
non-condensate atoms are moving with average velocity ${\bf v}_{s}$
relative to a stationary condensate. 
For example, the transformation (\ref{eq:gauge}) 
gives 
\ba
\tilde{g'}_{11}^{\down}(1,1')&=&e^{-i(\theta(1)-\theta(1'))}
\tilde{g}_{11}^{\down}(1,1') ,\nonumber \\
h'_{11}(1,1')&=&-i\sqrt{n_{c}(1)n_{c}(1')}.
\ea
One sees that, in the local rest frame (denoted by a prime), 
the order parameter $\Phi^{'}(\r,t)$ is real (see (\ref{eq:defh})). 

Equations (\ref{eq:tildeg2}) and (\ref{eq:tildeg3})  remain 
unchanged in form after this transformation to the local rest 
frame as long as $g_{0}^{-1}$ is replaced by 
(compare with (\ref{eq:godef}) in the lab frame):
\be
{g}_{0}^{-1}(1,1')=\left[ i {\bf \tau}_{3}
\frac{\partial}{\partial t_{1}}-\frac{\partial \theta(1)}
{\partial t_{1}}+\frac{1}{2}[\nabla_{1}+i{\bf \tau}_{3}
\nabla_{1}\theta(1) ]^{2}-U_{ext}(\r_{1})+\mu_{0} \right]
 \delta(1,1'). \label{eq:g0llf}
\ee
We recall that the superfluid velocity ${\bf v}_{s}\co$ 
and the local chemical potential $\mu_{c}¥\co$ are 
defined as the spatial and time derivatives of the phase
\cite{kk}, namely
\ba
m{\bf v}_{s}\co&\equiv& \nabla_{\bf R}\theta\co     
\nonumber \\
\frac{\partial \theta\co}{\partial T}
&\equiv&-\left[\mu_{c}¥\co-\mu_{0}
+\frac{1}{2}m {v_{s}}^{2}\co\right], 
\label{eq:defvsmu}
\ea
where, in the lab frame, the condensate wavefunction 
is given by $\Phi\co\equiv \sqrt{n_{c}\co}e^{i\theta\co}$.
Therefore, the gauge transformation changes the momentum 
${\bf p} \rightarrow {\bf p}- m{\bf v}_{s}$ as expected
for the momentum in the local rest frame
(when we Fourier transform, the gradient operator
 in (\ref{eq:g0llf}) becomes the momentum).
 
In Section II, we have written down the equations of motion 
(\ref{eq:tildeg2}) and (\ref{eq:tildeg3})
for the non-equilibrium real-time Green's functions. We 
now want to use these to derive a kinetic equation for the 
quasiparticle 
distribution function. We recall that (\ref{eq:tildeg2}) and 
(\ref{eq:tildeg3}) are matrix equations. In addition, 
(\ref{eq:tildeg2})  involves differential operators with 
the respect to the coordinates
($\r,t$) and (\ref{eq:tildeg3}) involves derivatives with 
respect to the coordinates ($\r',t'$). 
Since our single-particle Green's functions are functions of 
both coordinates (1,1$'$), one has to find a way to combine both 
equations 
to derive a single kinetic equation for a quasiparticle 
distribution function. We will discuss later how these 
Green's functions are related to the quasiparticle distribution 
function.
However, we emphasize that we will need to use both 
(\ref{eq:tildeg2}) 
and (\ref{eq:tildeg3})  to derive a kinetic equation for 
quasiparticles. 

In the Kadanoff-Baym procedure \cite{itg2,kb} 
one rewrites the equations of motion in 
terms of relative and center-of-mass 
space-time coordinates, defined by
\be
\r=\r_{1}-\r_{1'}, \hspace{5mm} t=t_{1}-t_{1'}; 
\hspace{5mm} {\bf R}=\frac{\r_{1}+\r_{1'}}{2},\hspace{5mm} 
T=\frac{t_{1}+t_{1'}}{2}.
\label{eq:defcom}
\ee
In the literature, this is sometimes known as the Wigner 
representation \cite{zubarevbook,huangjuho}. It 
allows one to separate variables describing ``slow'' and ``fast''
 processes in the system. In thermal equilibrium, 
 the Green's functions are only functions of 
the relative space-time coordinates $\r$ and $t$, 
and moreover are sharply peaked about $\r=0$ and $t=0$ \cite{kb}. 
Since we assume that the external disturbances are slowly 
varying in space and time 
(with a wavelength much larger than the thermal deBroglie 
wavelength), we expect that these slowly varying external 
disturbances will not change this dependence of 
$\tilde{g}^{\down}(\r,t;{\bf R},T)$ on small values of $\r$ and $t$.  
Therefore, our non-equilibrium correlation functions 
(like $\tilde{g}, \Sigma$, etc.) 
are assumed to be dominated by the small values of relative 
coordinates $(\r, t)$ (equivalently, by high momenta and frequencies 
in the Fourier transforms), but vary slowly as  
functions of the center-of-mass coordinates $\co$. 
 
 Using these key properties of the non-equilibrium correlation 
functions 
to simplify the equations, we now write  
(\ref{eq:tildeg2}) and (\ref{eq:tildeg3}) in terms of 
the  center-of-mass and relative coordinates 
\cite{itg2,kk,kt,itg1}, and take the trace of 
the resulting matrix equation to obtain: 
\ba 
&&\hat{\mathcal{L}}_{11}\tilde{g}_{11}^{<}(\r,t;{\bf R},T)
+\hat{\mathcal{L}}_{22}
\tilde{g}_{22}^{<}(\r,t;{\bf R},T)=
g\left[((\r \cdot \nabla_{\bf R}
+t\frac{\partial}{\partial T}) m)
\tilde{g}_{21}^{<}+\left(\left(\r \cdot \nabla_{\bf R}
+t\frac{\partial}{\partial T}\right) m^{*}
\right) \tilde{g}_{12}^{<}\right] \nonumber \\
&&+\int_{-\infty}^{\infty}d\bar{\r}d\bar{t}
 Tr\left(\Sigma^{>}(\r-\bar{\r},t-\bar{t})
 \hat{\tilde{g^{<}}}(\bar{\r},\bar{t})
-\Sigma^{<}(\r-\bar{\r},t-\bar{t})
\hat{\tilde{g^{>}}}(\bar{\r},\bar{t})\right).
\label{eq:popovcm} 
\ea 
As usual, for simplicity, the $({\bf R},T)$ dependence of the 
$\tilde{g},\Sigma$ and $m$ is left implicit.
The operators $\hat{{\mathcal L}}_{11}$ and 
$\hat{\mathcal{L}}_{22}$ on the left-hand side of
(\ref{eq:popovcm}) are defined as 
\ba
\hat{\mathcal{L}}_{11}&\equiv&i \frac{\partial}{\partial T}+(\r \cdot 
\nabla_{\bf R}+t\frac{\partial}{\partial T})\left(\mu \co 
-U_{eff}\co\right)+\frac{1}{m}\nabla_{\bf R}\cdot \nabla{\r}
 \nonumber \\
&+&i\left( (\r \cdot \nabla_{\bf R}+t\frac{\partial}{\partial T})
{\bf v}_{s}\co\right) \cdot \nabla_{\r}+i{\bf v}_{s}\co \cdot 
\nabla_{\bf R}
+i\nabla_{\bf R}\cdot {\bf v}_{s}\co  
\nonumber \\
\hat{\mathcal{L}}_{22}&\equiv& -i \frac{\partial}{\partial T}+
(\r \cdot \nabla_{\bf R}+t\frac{\partial}{\partial T})
\left(\mu \co -U_{eff}\co\right)
+\frac{1}{m}\nabla_{\bf R}\cdot \nabla{\r}\nonumber \\
&-& i\left( (\r \cdot \nabla_{\bf R}+t\frac{\partial}{\partial T})
{\bf v}_{s}\co\right) \cdot \nabla_{\r} 
-i{\bf v}_{s}\co \cdot \nabla_{\bf R}
-i\nabla_{\bf R}\cdot {\bf v}_{s}\co , 
\ea
where the effective dynamic HF field $U_{eff}\co$ is given by
\be
U_{eff}\co=U_{ext}({\bf R})+2g(n_{c}\co +\tilde{n}\co).
\ee
We emphasize that in the expansion for the small values of relative 
coordinates $(\r,t)$, we did not keep all terms of order 
$\partial/\partial T$ and $\nabla_{\bf R}$ in (\ref{eq:popovcm}). 
These additional terms that we have neglected 
contribute to the many-body renormalization effects, i.e. how the 
two-particle interaction changes the dispersion relation 
of the quasiparticles due to terms of second order in $g$. 
Such corrections involve the 
real part of the second-order Beliaev self-energies. 
The Bogoliubov-Popov quasiparticle 
approximation we use for the spectral densities $a_{\alpha \beta}$
in (\ref{eq:defagamma}) do not include such second-order effects. 
 In the present paper, we concentrate on the damping effects 
 associated with  the collisional self-energies 
 $\Sigma^{\down}$ on
the right-hand side of (\ref{eq:popovcm}).
For further discussion of the KB formalism related to going 
past the simple Bogoliubov quasiparticle approximation, 
see Ch. 9 of Ref. \cite{kb} and Ch.6 in 
Ref. \cite{zubarevbook}. 

The double Fourier transform of (\ref{eq:popovcm}) 
gives
\ba 
&&\hat{\mathcal{L}}_{11}\tilde{g}_{11}^{<}+\hat{\mathcal{L}}_{22}
\tilde{g}_{22}^{<}-g\nabla_{\bf R}m^{*}\cdot 
\nabla{\vp}\tilde{g}_{12}^{<}
-g\nabla_{\bf R}m\cdot \nabla{\vp}\tilde{g}_{21}^{<} 
+g\frac{\partial m^{*}}{\partial T} \frac{\partial \tilde{g}_{12}^{<}}
{\partial \omega}+g\frac{\partial m}{\partial T} 
\frac{\partial \tilde{g}_{21}^{<}}{\partial \omega}
\nonumber \\
&&=Tr\left(\hat{\Sigma}^{>}\pomrt \hat{\tilde{g^{<}}}\pomrt
-\hat{\Sigma}^{<}\pomrt \hat{\tilde{g^{>}}}\pomrt \right),  
\label{eq:kineqquasi} 
\ea 
 with $\tilde{g}^{<}_{\alpha \beta}\equiv 
\tilde{g}^{<}_{\alpha \beta}\pomrt$ and 
\ba
\hat{\mathcal{L}}_{11}&=&\frac{\partial}{\partial T}+
\nabla_{\bf p}\left[\nee+\vp\cdot {\bf v}_{s}\right]\cdot 
\nabla_{\bf R} - \nabla_{\bf R}\left[\nee
+\vp\cdot {\bf v}_{s}\right] 
\cdot \nabla_ \vp +\frac{\partial}{\partial T}
\left[\nee+\vp \cdot {\bf v}_{s} \right] 
\frac{\partial}{\partial\omega}
\nonumber \\
\hat{\mathcal{L}}_{22}&=&
-\frac{\partial}{\partial T}
+\nabla_{\bf p}\left[\nee-\vp\cdot {\bf v}_{s}\right]\cdot 
\nabla_{\bf R} - \nabla_{\bf R}\left[\nee
-\vp\cdot {\bf v}_{s}\right] 
\cdot \nabla_ \vp +\frac{\partial}{\partial T}
\left[\nee-\vp \cdot {\bf v}_{s} \right] 
\frac{\partial}{\partial\omega}.
\ea
The result in (\ref{eq:kineqquasi}) gives an equation closely related 
to the quasiparticle kinetic equation we are trying to derive.
Here, $\nee$ is defined by
\be
\nee \co=\frac{p^{2}}{2m}+U_{ext}({\bf R})+2gn\co-\mu_{c}\co.
\label{eq:defspe}
\ee
In Section VI, we shall see that the condensate 
chemical potential $\mu_{c}\co$ is given by
\be
\mu_{c}=-\frac{\nabla_{\bf R}\sqrt{n_{c}\co}}{2m\sqrt{n_{c}\co}}
+U_{ext}({\bf R})+U\co+g\left[2\tilde{n}\co+n_{c}\co\right].
\label{eq:chemicalpot}
\ee

A kinetic equation for thermally excited atoms in a trapped Bose gas 
can be written in terms of distribution 
functions for either atoms or for 
quasiparticle excitations. In our earlier papers 
\cite{itg2,itg1}, we transformed the equations of motion for a 
non-equilibrium Green's functions at high temperatures 
into a kinetic equation for a 
single-particle distribution function $f(\vp,{\bf R},T)$ 
describing the non-condensate atoms. The latter are assumed to have
a Hartree-Fock spectrum. 
If one wants to use a more realistic spectrum valid at low 
temperatures, it is much more convenient to work within a 
quasiparticle 
picture. In the theory of 
Bose-condensed trapped gases, one introduces quasiparticles
by expressing the quantum field operators for the non-condensate 
atoms as a coherent superposition of creation and annihilation 
operators 
for Bose quasiparticles, with the weights given by 
the usual Bose-coherence factors $u$ and $v$
\be
\tilde{\psi}({\bf R},T) \equiv \sum_{i}\left[u_{i}({\bf R})
\hat{\alpha}_{i}e^{-iE_{i}T/\hbar}+v^{*}_{i}({\bf R})
\hat{\alpha}^{\dag}_{i}e^{iE_{i}T/\hbar}\right].
\label{eq:quasitransf}
\ee
In the semiclassical approximation, (\ref{eq:quasitransf})
becomes
\be
\tilde{\psi}({\bf R},T) \equiv \int\frac{d\vp}{(2\pi)^{3}}
\left[u_{p}({\bf R})
\hat{\alpha}_{p}e^{-iE_{p}T/\hbar}+v^{*}_{p}({\bf R})
\hat{\alpha}^{\dag}_{p}e^{iE_{p}T/\hbar}\right].
\ee
Here $\hat{\alpha}^{\dag}_{p}$ and $\hat{\alpha}_{p}$ are 
the Bogoliubov quasiparticle 
creation and annihilation operators, respectively, which obey the 
usual Bose commutation relations. One can see that 
creating an atom with momentum $\vp$ is equivalent 
to creating a quasiparticle with momentum $\vp$ with amplitude 
$u_{p}$ 
and at the same time, destroying a quasiparticle
with momentum $-\vp$ and amplitude 
$v_{p}$. The quasiparticle distribution function is given 
by the statistical average of the 
quasiparticle operators, i.e., $f(\vp)\equiv \langle 
\hat{\alpha}^{\dag}_{p}\hat{\alpha}_{p}\rangle$. 
We  recall \cite{itg2,itg1} that  
distribution function for atoms $f_{at}(\vp,\omega;{\bf R},T)$ 
is directly related 
to the diagonal Green's function, namely
\be
f_{at}(\vp,\omega;{\bf R},T)=-i
\tilde{g}_{11}^{<}(\vp,\omega;{\bf R},T).
\label{eq:atomf}
\ee 
In terms of quantum field operators, $\tilde{g}_{11}^{<}$ is 
given by (see Ch. 9 of Ref.\cite{kb})
\be
\tilde{g}_{11}^{<}(\vp,\omega;{\bf R},T)=i\int d\r dt 
e^{-i\vp \cdot \r+i\omega t}\langle \tilde{\psi}^{\dag}
({\bf R}-\frac{\r}{2},T-\frac{t}{2})
\tilde{\psi}({\bf R}+\frac{\r}{2},T+\frac{t}{2})\rangle. 
\ee

The usual Boltzmann equation is expressed in terms of the Wigner 
distribution function $f_{W}(\vp,{\bf R},T)$ \cite{itg2}. 
This limits the description to the 
semiclassical approximation because it is assumed that the position 
and momentum of the particles can be defined simultaneously. In order 
to use this kind of distribution function for quantum systems, it is 
necessary 
to perform some type of averaging in order to remove effects due to 
the uncertainty principle. In our work, we want to derive a kinetic 
equation for the quasiparticles which is valid at all temperatures 
and 
therefore the semiclassical 
approximation will no longer be valid. To include the quantum 
effects, we introduce the {\it quasiparticle} 
distribution function $f\pomrt$ with an additional variable $\omega$ 
in the following way  \cite{kk,zubarevbook,huangjuho,mahanbook}
\ba 
&&\tilde{g}_{\alpha \beta}^{<}\pomrt \equiv i 
 a_{\alpha \beta}\pomrt f \pomrt \nonumber \\ 
&&\tilde{g}_{\alpha \beta}^{>}\pomrt 
\equiv i  a_{\alpha \beta}\pomrt [1+f \pomrt], 
\label{eq:deff}
\ea 
where the spectral density $a_{\alpha \beta}$ is defined in 
(\ref{eq:defagamma}). Using the Bogoliubov-Popov approximation
for the spectral density in (\ref{eq:sdbp}), one can use  
(\ref{eq:deff}) to obtain the well-known relation between 
the quasiparticle distribution function $f\pomrt$ and the 
atom distribution function $f_{at}$ in (\ref{eq:atomf}), namely
\be
f_{at}\pomrt
=\left(u_{p}^{2}({\bf R},T)+v_{p}^{2}({\bf R},T)
 \right) f\pomrt +v_{p}^{2}({\bf R},T).
\label{eq:parttoquasi}
\ee
The semiclassical Wigner distribution function is obtained by taking 
the frequency integral of the atom distribution function 
defined in (\ref{eq:parttoquasi}). 

One can show, using (\ref{eq:tildeg2}) and (\ref{eq:tildeg3}), that 
the spectral function $a_{\alpha \beta}$ in  (\ref{eq:defagamma})
satisfies
\ba
\hat{\mathcal{L}}_{11}a_{11}+\hat{\mathcal{L}}_{22}
a_{22}-g\nabla_{\bf R}m^{*}\cdot \nabla_{\vp}a_{12}
-g\nabla_{\bf R}m\cdot \nabla_{\vp}a_{21}
-g\frac{\partial m^{*}}{\partial T} 
\frac{\partial a_{12}}{\partial \omega}
-g\frac{\partial m}{\partial T} 
\frac{\partial a_{21}}{\partial \omega}=0.
\label{eq:eqfora}
\ea
Using (\ref{eq:deff}) and (\ref{eq:eqfora}), one can 
rewrite the kinetic equation 
(\ref{eq:kineqquasi}) for $\tilde{g}_{\up}$ 
to obtain a new kinetic equation specifically for the 
quasiparticle distribution function $f (\vp,\omega;{\bf R},T)$
 in the following form:
\ba
&&a_{11}\hat{\mathcal{L}}_{11}f+a_{22}\hat{\mathcal{L}}_{22}
f-a_{12}g\nabla_{\bf R}m^{*}\cdot \nabla_{\vp}f-a_{21}
g\nabla_{\bf R}m \cdot \nabla_{\vp}f \nonumber \\
&+&a_{12}\frac{\partial m^{*}}{\partial T}
\frac{\partial f}{\partial \omega}+a_{21}\frac{\partial m}{\partial T}
\frac{\partial f}{\partial 
\omega}=f Tr(\Sigma^{>}\hat{a})-(1+f)Tr(\Sigma^{<}\hat{a}).  
\label{eq:generalke}
\ea  
We notice that (\ref{eq:generalke}) includes  terms involving
 $\partial/\partial \omega$. The additional variable $\omega$ 
in $f(\vp,\omega;{\bf R},T)$ results in new streaming terms on the 
left side of (\ref{eq:generalke}). These terms are not present in 
the semiclassical kinetic equation (which is obtained from 
(\ref{eq:generalke}) 
integrating over $\omega$), which shows that the terms 
involving  $\partial/\partial \omega$ are of quantum origin
\cite{zubarevbook,huangjuho,mahanbook}.  

Eq. (\ref{eq:generalke}) is the most general form for a 
kinetic equation for the quasiparticle distribution 
function $f$ within our model. To derive (\ref{eq:generalke}), 
we have only assumed that the external disturbances vary slowly in 
space and time, and therefore all relevant physical quantities  
vary slowly as function of center-of-mass coordinates 
$({\bf R},T)$ defined in (\ref {eq:defcom}). 
The other assumption that we made is that one can 
introduce a quasiparticle distribution function $f$ through the 
definition in
(\ref{eq:deff}). Of course, at this stage, one could say that we are 
only replacing one unknown function with another.
The generalized kinetic equation for a Bose-condensed system 
(\ref{eq:generalke}) was first derived by Kane \cite{kt}.
From (\ref{eq:generalke}), we see that the 
general structure of the collision integral $I$ has 
the following form \cite{kk,kt}
\be
I[f(\vp,{\bf R},T)]\equiv \int \frac{d\omega}{2\pi}\left[f 
Tr(\hat{\Sigma}^{>}\hat{a})-
(1+f)Tr(\hat{\Sigma}^{<}\hat{a}) \right].
\label{eq:generalcoll}
\ee
Using (\ref{eq:deff}) in the the general expression for 
the non-equilibrium Beliaev self-energies $\Sigma^{\down}$ given in
(\ref{eq:sigmabeliaev}), one can prove that the collision integral 
given by (\ref{eq:generalcoll}) conserves momentum
(see Appendix for details),
\be
\int d\vp \vp I[f(\vp,{\bf R},T)]=0.
\label{eq:conservedp}
\ee
One can also prove that (\ref{eq:generalcoll})  
conserves energy as well. To prove this, however, 
we need to work within a specific 
approximation for the single-particle 
spectral density $ a_{\alpha \beta}\pomrt$. 

\section{Kinetic equation in the Bogoliubov-Popov approximation}

In this section, we use the results of Section III to 
derive a quasiparticle kinetic equation within
the Bogoliubov-Popov approximation.
More precisely, this means that we will
use the spectral densities with the Bogoliubov-Popov 
quasiparticle excitation energies \cite{kb,kt,fw}
\ba
&&a_{11}\pomrt=2\pi[u_{p}^{2}\delta(\omega-{\bf v}_{s}\cdot 
\vp-E_{p})-v_{p}^{2}
\delta(\omega-{\bf v}_{s}\cdot \vp+E_{p})] \nonumber \\
&&a_{12}\pomrt=-2\pi u_{p}v_{p}[\delta(\omega-{\bf v}_{s}
\cdot \vp-E_{p})-\delta(\omega-{\bf v}_{s}\cdot \vp+E_{p})] 
\nonumber \\
&&a_{21}\pomrt = a_{12}\pomrt  \nonumber \\
&& a_{22}\pomrt =-a_{11}(-\vp,-\omega;{\bf R},T).
\label{eq:sdbp}
\ea
Here, the Bose-coherence factors $u({\bf R},T)$ and 
$v({\bf R},T)$ are given by \cite{fw}
\be
u_{p}^{2}\co =\frac{\nee \co+E_{p}\co}{2E_{p}\co}, \hspace{5mm}
u_{p}^{2}-v_{p}^{2}=1, \hspace{5mm} 
u_{p}v_{p}=\frac{gn_{c}\co}{2E_{p}\co}
\label{eq:defuv}
\ee 
and the quasiparticle energy $E_{p}$ is given by 
\be 
E_{p}\co =\sqrt{\nee^{2}\co -(gn_{c}\co)^{2}}.
\ee
We emphasize that the spectral densities in (\ref{eq:sdbp}) could be 
derived in the quasiparticle approximation from the general equations 
of motion for the Green's functions as it has been shown in 
\cite{kt}. We simply start with them as input into our general 
formalism.

In the Thomas-Fermi approximation \cite{strtheory},
one neglects the quantum pressure 
term in (\ref{eq:chemicalpot}), in which case the quasiparticle 
energy 
$E_{p}$ reduces to the usual Bogoliubov excitation energy
\be
E_{p}\co=\sqrt{\epsilon_{p}^{2}+2gn_{c}\co \epsilon_{p}},  
\label{eq:defquasie}
\ee
where $\epsilon_{p}=p^{2}/2m$. We note that spectral densities 
$a_{\alpha \beta}(\vp,\omega;{\bf R},T)$ in (\ref{eq:sdbp}) 
exhibit both positive and 
negative energy poles. In the Hartree-Fock approximation used in our 
earlier work \cite{itg2}, $u_{p}^{2}=1$ and $v_{p}^{2}=0$. 
Physically, (\ref{eq:sdbp})
corresponds to the assumption that the thermal cloud 
can be considered as a gas of 
weakly-interacting single-particle excitations with the excitation 
energy given by (\ref {eq:defquasie}). One can check 
explicitly that (\ref{eq:sdbp}) do satisfy the general equation 
of motion given in (\ref{eq:eqfora}). Note that in the 
literature, $v_{p}$ is sometimes defined with the opposite sign, such 
that $u_{p}v_{p}$ in (\ref{eq:defuv}) is negative.

Substituting the spectral densities (\ref{eq:sdbp})  into 
(\ref{eq:generalke}), we can now derive a kinetic equation 
for the quasiparticles. After lengthy algebra, we obtain    
\ba
&&\int_{-\infty}^{\infty}\frac{d\omega}{2\pi}
\delta(\omega-{\bf v}_{s}\cdot \vp-E_{p})\left[ \frac{\partial 
f}{\partial T} +\nabla_{\vp}\left(E_{p}+{\bf v}_{s}\cdot 
\vp\right)\cdot \nabla_{\bf R}f-\nabla_{\bf R}\left(E_{p}+{\bf 
v}_{s}\cdot 
\vp\right)\cdot \nabla_{\vp}f \right. \nonumber \\
&+&\frac{\partial}{\partial T} \left(E_{p}+{\bf v}_{s}\cdot 
\vp\right) \frac{\partial f}{\partial \omega}+
u_{p}^{2}\left( (1+f)\Sigma_{11}^{<}-f\Sigma_{11}^{>}\right)
+v_{p}^{2}\left( (1+f)\Sigma_{22}^{<}-f\Sigma_{22}^{>}\right)
\nonumber \\
&-&\left. u_{p}v_{p}\left( (1+f)(\Sigma_{12}^{<}+\Sigma_{21}^{<})-
f(\Sigma_{12}^{>}+\Sigma_{21}^{>})\right) \right]
\nonumber \\
&+&\int_{-\infty}^{\infty}\frac{d\omega}{2\pi}
\delta(\omega-{\bf v}_{s}\cdot \vp+E_{p})\left[ \frac{\partial 
f}{\partial T} +\nabla_{\vp}\left(-E_{p}+{\bf v}_{s}\cdot 
\vp\right)\cdot \nabla_{\bf R}f-\nabla_{\bf R}\left(-E_{p}+{\bf 
v}_{s}\cdot 
\vp\right)\cdot \nabla_{\vp}f \right. \nonumber \\
&+&\frac{\partial}{\partial T} \left(-E_{p}+{\bf v}_{s}\cdot 
\vp\right) \frac{\partial f}{\partial \omega}+
v_{p}^{2}\left( (1+f)\Sigma_{11}^{<}-f\Sigma_{11}^{>}\right)
+u_{p}^{2}\left( (1+f)\Sigma_{22}^{<}-f\Sigma_{22}^{>}\right)
\nonumber \\
&-&\left. u_{p}v_{p}\left( (1+f)(\Sigma_{12}^{<}+\Sigma_{21}^{<})-
f(\Sigma_{12}^{>}+\Sigma_{21}^{>})\right) \right]=0,
\label{eq:completeke}
\ea
where $\Sigma_{\alpha \beta}=\Sigma_{\alpha \beta}\pomrt$.
This is a kinetic equation for the frequency dependent 
quasiparticle distribution function $f$ expressed 
in terms of an integral over 
both positive and negative energy poles. 
If we recall the expression for Bose coherence factors 
$u$ and $v$ given by (\ref {eq:defuv}), we note
that the second term in (\ref{eq:completeke}) is 
the same as the first term in 
(\ref{eq:completeke}) if we replace $-E_{p}$ with $E_{p}$. Therefore, 
it is sufficient only to consider the first term   
to obtain the kinetic equation for the quasiparticle distribution 
function defined by
\be
f_{qp}(\vp,{\bf R},T) \equiv 
f(\vp, \omega-{\bf v}_{s}\cdot\vp=E_{p};{\bf R},T).
\label{eq:quasif}
\ee
We obtain finally
\ba
\left[\frac{\partial f_{qp}¥}{\partial T}\right. &+&\left.
\nabla_{\vp}\left(E_{p}
+{\bf v}_{s}\cdot \vp\right)\cdot \nabla_{\bf R}f_{qp}
-\nabla_{\bf R}\left(E_{p}+{\bf v}_{s}\cdot\vp\right)
\cdot \nabla_{\vp}f_{qp} \right]=I[f_{qp}]. 
\label{eq:ppoles}
\ea
Here the collision integral $I[f_{qp}(p,{\bf R},T)]$, 
defined in (\ref{eq:generalcoll}),
becomes
\ba
I[f_{qp}(p,{\bf R},T)]&\equiv &\int_{-\infty}^{\infty}
\frac{d\omega}{2\pi}
\left[u_{p}^{2}\left( (1+f_{qp})\Sigma_{11}^{<}-f_{qp}
\Sigma_{11}^{>}\right)+v_{p}^{2}\left( (1+f_{qp})
\Sigma_{22}^{<}-f_{qp}¥\Sigma_{22}^{>}\right)
\right. \nonumber \\
&+&\left. u_{p}v_{p}\left( (1+f_{qp})
(\Sigma_{12}^{<}+\Sigma_{21}^{<})-
f_{qp}(\Sigma_{12}^{>}+\Sigma_{21}^{>})\right) 
\right].
\label{eq:collintegralgeneral}
\ea

To evaluate the collision integral in (\ref{eq:collintegralgeneral}),
we need to 
choose a specific approximation for the second-order 
self-energy $\Sigma_{\alpha \beta}$. 
Here, we use the second order Beliaev approximation given by 
(\ref{eq:sigmabeliaev}), the Fourier transform of 
which is 
\ba
&&\hat{\Sigma}^{\down}({\bf p},\omega;{\bf R},T)
=-\frac{1}{2}g^{2}\int \frac{d{\bf p}_{i}d\omega_{i}}
{(2\pi)^{8}}\delta(\omega+\omega_{1}
-\omega_{2}-\omega_{3})\delta({\bf p}
+{\bf p}_{1}-{\bf p}_{2}-{\bf p}_{3}) 
\nonumber \\
&\times &\left[ \hat{\tilde{g}}^{\down}
({\bf p}_{2},\omega_{2})Tr\left[\hat{\tilde{g}}^{\up}
({\bf p}_{1},\omega_{1})\hat{\tilde{g}}^{\down}
({\bf p}_{3},\omega_{3})\right]
+2\hat{\tilde{g}}^{\down}({\bf p}_{2},\omega_{2})
\hat{\tilde{g}}^{\up}({\bf p}_{1},\omega_{1})
\hat{\tilde{g}}^{\down}({\bf p}_{3},\omega_{3})
\right.\nonumber \\
&+&\hat{\tilde{g}}^{\down}({\bf p}_{2},\omega_{2})
Tr\left[\hat{h}({\bf p}_{1},\omega_{1})
\hat{\tilde{g}}^{\down}({\bf p}_{3},\omega_{3})
+\hat{\tilde{g}}^{\up}({\bf p}_{1},\omega_{1})\hat{h}
({\bf p}_{3},\omega_{3})\right]\nonumber \\
&+&\hat{h}({\bf p}_{2},\omega_{2})Tr\left[\hat{\tilde{g}}^{\up}
({\bf p}_{1},\omega_{1})\hat{\tilde{g}}^{\down}
({\bf p}_{3},\omega_{3})\right] 
+2\hat{h}({\bf p}_{2},\omega_{2})\hat{\tilde{g}}^{\up}
({\bf p}_{1},\omega_{1})\hat{\tilde{g}}^{\down}
({\bf p}_{3},\omega_{3}) \nonumber \\
&+&\left.2\hat{\tilde{g}}^{\down}({\bf p}_{2},\omega_{2})
\left[\hat{h}({\bf p}_{1},\omega_{1})
\hat{\tilde{g}}^{\down}({\bf p}_{3},\omega_{3})
+\hat{\tilde{g}}^{\up}({\bf p}_{1},\omega_{1})\hat{h}
({\bf p}_{3},\omega_{3})\right]
\right].
\label{eq:sigmabeliaevft}
\ea
As usual, the $({\bf R},T)$ dependence of the functions 
$\Sigma, \tilde{g}$ and $h$ on the right-hand side has 
been suppressed for simplicity of notation. 
The quasiparticle energy $E_{p}({\bf R},T)$ 
in (\ref{eq:ppoles}) is the energy of the quasiparticles in the local 
rest frame (${\bf v}_{s}=0$). The Beliaev second-order expression  
(\ref{eq:sigmabeliaevft}) consists 
of two kinds of contributions: (1) Terms that include 
both the condensate propagator
 $h$ {\it and} the non-condensate propagators $\tilde{g}$; 
 (2) Terms that include the non-condensate 
propagators $\tilde{g}$ only. The first type of contribution 
will give rise to the collision integral that describe 
collisions that include one condensate atom interacting with 
the thermally excited quasiparticles. 
As in earlier work  \cite{zngjltp,itg2,kd}, 
we denote this part of the collision integral as $C_{12}$, 
indicating that we go from 
1 thermally excited quasiparticle (and one condensate atom) 
to 2 thermally excited quasiparticles. The second type of 
contribution in  (\ref{eq:sigmabeliaevft}) only 
includes  non-condensate propagators. We denote 
this part of the collision integral as $C_{22}$, indicating that it 
describes collisions where 2 thermally excited quasiparticles are 
scattered into 2 excited quasiparticles. At low 
temperatures, when the number of thermally excited 
quasiparticles is small, we can neglect the $C_{22}$ 
collision integral relative to $C_{12}$. 

From (\ref{eq:conservedp}), it follows that both $C_{12}$ and $C_{22}$
conserve momentum (see Appendix for details), i.e., 
\ba
\int d\vp \vp C_{12}=0 \nonumber \\
\int d\vp \vp C_{22}=0.
\ea
In addition, using (\ref{eq:sdbp}) in (\ref{eq:sigmabeliaevft}) one 
can show that the collision integral in (\ref {eq:generalcoll}) 
conserves 
quasiparticle energy $E_{p}$ and therefore 
both $C_{12}$ and $C_{22}$ will satisfy the conditions
\ba
\int d\vp E_{p} C_{12}=0 \nonumber \\
\int d\vp E_{p} C_{22}=0.
\ea

 One can show that for slowly varying external disturbances,  
the condensate propagator in (\ref{eq:defh}) can be approximated by
\be
\hat{h}\pomrt =n_{c}({\bf 
R},T)(2\pi)^{4}\delta(p)\delta(\omega)
\left( \begin{array}{cc} 1 \hspace*{3mm} 1\\ 
1 \hspace*{3mm} 1 
\end{array}\right).
\label{eq:localcond}
\ee 
To evaluate the collision integral $C_{12}$  
in terms of the Bose coherence factors {\it u} and {\it v},
 we need to substitute 
(\ref {eq:sdbp}) and (\ref{eq:localcond}) 
into (\ref{eq:sigmabeliaevft}). One can simplify 
(\ref{eq:sigmabeliaevft}) greatly using the following exact symmetry 
relations;
\ba
&&\tilde{g}_{22}^{\up}\pomrt=\tilde{g}_{11}^{\down}
(-\vp,-\omega;{\bf R},T),
\nonumber \\
&&\tilde{g}_{12}^{\up}\pomrt=\tilde{g}_{12}^{\down}
(-\vp,-\omega;{\bf R},T),
\nonumber \\
&&\tilde{g}_{21}^{\up}\pomrt=\tilde{g}_{21}^{\down}
(-\vp,-\omega;{\bf R},T).
\label{eq:propertiesofg}
\ea
After some algebra, one finds the following expressions 
for the non-equilibrium self-energy  (considering now 
only the $C_{12}$ terms, which include one condensate propagator):
\ba
&&\Sigma_{11}^{\down}(\vp,\omega;{\bf R},T)=-g^{2}\int \frac{d\vp_{2}
d\omega_{2}}{(2\pi)^{4}}
n_{c}\co\left[ 2\tilde{g}_{11}^{\down}(\vp_{2},\omega_{2})
\tilde{g}_{11}^{\down}(\vp-\vp_{2},\omega-\omega_{2})
\right. \nonumber \\
&+&4\tilde{g}_{11}^{\down}(\vp_{2},\omega_{2})
\tilde{g}_{11}^{\up}(\vp_{2}-\vp,\omega_{2}-\omega)
+8\tilde{g}_{12}^{\down}(\vp_{2},\omega_{2})
\tilde{g}_{11}^{\down}(\vp-\vp_{2},\omega-\omega_{2})
\nonumber \\
&+&\left.4\tilde{g}_{12}^{\down}(\vp_{2},\omega_{2})
\tilde{g}_{12}^{\down}(\vp-\vp_{2},\omega-\omega_{2})
\right],
\label{eq:beliaev11}
\ea
\ba
&&\Sigma_{22}^{\down}(\vp,\omega;{\bf R},T)=-g^{2}\int 
\frac{d\vp_{2}d\omega_{2}¥}{(2\pi)^{4}¥}
n_{c}\co\left[ 4\tilde{g}_{12}^{\down}(\vp_{2},\omega_{2})
\tilde{g}_{12}^{\down}(\vp-\vp_{2},\omega-\omega_{2})
\right. \nonumber \\
&+&4 \tilde{g}_{11}^{\down}(\vp_{2},\omega_{2})
\tilde{g}_{11}^{\up}(\vp_{2}-\vp,\omega_{2}-\omega)
+8\tilde{g}_{12}^{\down}(\vp_{2},\omega_{2})
\tilde{g}_{11}^{\up}(\vp_{2}-\vp,\omega_{2}-\omega)
\nonumber \\
&+&\left. 2\tilde{g}_{11}^{\up}(-\vp_{2},-\omega_{2})
\tilde{g}_{11}^{\up}(\vp_{2}-\vp,\omega_{2}-\omega)
\right], 
\label{eq:beliaev22}
\ea
\ba
&&\Sigma_{12}^{\down}¥(\vp,\omega;{\bf R},T)=-g^{2}
\int \frac{d\vp_{2}d\omega_{2}}{(2\pi)^{4}}
n_{c}\co\left[6\tilde{g}_{12}^{\down}(\vp_{2},\omega_{2})
\tilde{g}_{12}^{\down}(\vp-\vp_{2},\omega-\omega_{2}) 
\right. \nonumber \\
&+&4\tilde{g}_{12}^{\down}(\vp_{2},\omega_{2})
\tilde{g}_{11}^{\down}(\vp-\vp_{2},\omega-\omega_{2})
+4\tilde{g}_{11}^{\down}(\vp_{2},\omega_{2})
\tilde{g}_{11}^{\up}(\vp_{2}-\vp,\omega_{2}-\omega)
\nonumber \\
&+&\left.4\tilde{g}_{12}^{\down}(\vp_{2},\omega_{2})
\tilde{g}_{11}^{\up}(\vp_{2}-\vp,\omega_{2}-\omega)
\right], 
\label{eq:beliaev12}
\ea
\ba
&&\Sigma_{21}^{\down}(\vp,\omega;{\bf R},T)
=\Sigma_{12}^{\down}(\vp,\omega;{\bf R},T).
\ea
Again, the $({\bf R},T)$ dependence of the $\tilde{g}$'s is suppressed
 on the right-hand side.
One notices that these expressions have the same structure as the 
thermal equilibrium ones obtained by Shi and Griffin, 
as well as others \cite{shi,giorgini99,fedichev}. This is expected 
since 
the whole structure of our theory is only valid for a systems 
slightly perturbed from thermal equilibrium,
with all physical quantities assumed to vary slowly 
as functions of center-of-mass coordinates ({\bf R},$T$). 
The entire KB formalism reduces to the usual equilibrium 
self-energies in the appropriate limit, which is one of its strengths.

One can show using the general properties of the non-equilibrium 
Green's functions in (\ref{eq:propertiesofg}) in conjunction with
(\ref{eq:deff}) that the 
quasiparticle distribution function $f\pomrt$ satisfies  
the exact relation
\be
f(-\vp,-\omega;{\bf R},T)=-(1+f(\vp,\omega;{\bf R},T)).
\label{eq:minusf}
\ee
If we introduce following standard abbreviations for the 
Bose-coherence factors 
\be
A_{p}\equiv u_{p}^{2}, \hspace{5mm} B_{p}\equiv v_{p}^{2},
\hspace{5mm} C_{p}\equiv -u_{p}v_{p},  
\ee
the self-energies in (\ref{eq:beliaev11})-(\ref{eq:beliaev12}) can be 
written as ( using (\ref{eq:minusf}))
\ba
\Sigma_{11}^{\down}(\vp,\omega;{\bf R},T)&=&g^{2}\int \frac{d\vp_{2}
d\omega_{2}}{(2\pi)^{2}} n_{c}\co \left( {\begin{array}{c}  
(1+f_{1})(1+f_{2}) \\ f_{1} f_{2} \end{array}} \right)
\nonumber \\
&&\left[\left( 2A_{1}A_{2}+8A_{1}C_{2}
+4C_{1}C_{2}+4B_{1}A_{2}\right)
\delta(\omega_{2}-E_{2})\delta(\omega_{1}-E_{1}) 
\right. \nonumber \\
&-&\left( 2B_{1}A_{2}+8B_{1}C_{2}+4C_{1}C_{2}+4A_{1}A_{2}\right)
\delta(\omega_{2}-E_{2})\delta(\omega_{1}+E_{1}) 
\nonumber \\
&-&\left( 2A_{1}B_{2}+8A_{1}C_{2}+4C_{1}C_{2}+4B_{1}B_{2}\right)
\delta(\omega_{2}+E_{2})\delta(\omega_{1}-E_{1})
\nonumber \\
&+& \left.\left( 
2B_{1}B_{2}+8B_{1}C_{2}+4C_{1}C_{2}+4A_{1}B_{2}\right)
\delta(\omega_{2}+E_{2})\delta(\omega_{1}+E_{1}) \right],
\ea  
\ba
\Sigma_{12}^{\down}(\vp,\omega;{\bf R},T)&=&g^{2}\int \frac{d\vp_{2}
d\omega_{2}}{(2\pi)^{2}} n_{c}\co \left( {\begin{array}{c}  
(1+f_{1})(1+f_{2}) \\ f_{1} f_{2} \end{array}} \right)
\nonumber \\
&&\left[ \left(6C_{1}C_{2}+4A_{1}C_{2}+4B_{1}A_{2}+4B_{1}C_{2}\right)
\delta(\omega_{2}-E_{2})\delta(\omega_{1}-E_{1}) \right. 
\nonumber \\
&-&\left( 4B_{1}C_{2}+4A_{1}C_{2}+6C_{1}C_{2}+4A_{1}A_{2}\right)
\delta(\omega_{2}-E_{2})\delta(\omega_{1}+E_{1}) 
\nonumber \\
&-&\left(4A_{1}C_{2}+4B_{1}C_{2}+6C_{1}C_{2}+4B_{1}B_{2}\right)
\delta(\omega_{2}+E_{2})\delta(\omega_{1}-E_{1})
\nonumber \\
&+&\left.\left( 4A_{1}B_{2}+4B_{1}C_{2}+6C_{1}C_{2}+4A_{1}C_{2}\right)
\delta(\omega_{2}+E_{2})\delta(\omega_{1}+E_{1})
\right],
\ea  
and
\ba
\Sigma_{22}^{\down}(\vp,\omega;{\bf R},T)&=&g^{2}\int \frac{d\vp_{2}
d\omega_{2}}{(2\pi)^{2}} n_{c}\co \left( {\begin{array}{c}  
(1+f_{1})(1+f_{2}) \\ f_{1} f_{2} \end{array}} \right)
\nonumber \\
&&\left[\left( 2B_{1}B_{2}+8B_{1}C_{2}+4C_{1}C_{2}+4B_{1}A_{2}\right)
\delta(\omega_{2}-E_{2})\delta(\omega_{1}-E_{1}) \right. 
\nonumber \\
&-&\left( 2A_{1}B_{2}+8A_{1}C_{2}+4C_{1}C_{2}+4A_{1}A_{2}\right)
\delta(\omega_{2}-E_{2})\delta(\omega_{1}+E_{1}) 
\nonumber \\
&-&\left( 2B_{1}A_{2}+8B_{1}C_{2}+4C_{1}C_{2}+4B_{1}B_{2}\right)
\delta(\omega_{2}+E_{2})\delta(\omega_{1}-E_{1})
\nonumber \\
&+&\left.\left( 2A_{1}A_{2}+8A_{1}C_{2}+4C_{1}C_{2}+4A_{1}B_{2}\right)
\delta(\omega_{2}+E_{2})\delta(\omega_{1}+E_{1})
\right],
\ea  
where $\vp-\vp_{2}\equiv \vp_{1}$ and 
$\omega-\omega_{2}\equiv \omega_{1}$.

Using these results, we can finally evaluate 
the $C_{12}$ collision integral given in 
(\ref{eq:collintegralgeneral}) 
\ba
&&C_{12}[f]=2g^{2}n_{c}\co\int \frac{d\vp_{1}d\vp_{2}}{(2\pi)^{2}}
\left[(1+f)f_{1}f_{2}-f(1+f_{1})(1+f_{2})\right]
\delta(\vp-\vp_{1}-\vp_{2})\nonumber \\
&&\left[ \left( (u_{1}-v_{1})(u_{p}u_{2}+v_{p}v_{2})+
(u_{2}-v_{2})(u_{p}u_{1}+v_{p}v_{1})
-(u_{p}-v_{p})(u_{1}v_{2}+v_{1}u_{2}) \right)^{2}
\delta(E_{p}-E_{1}-E_{2})\right.\nonumber \\
&+&2 \left( (u_{1}-v_{1})(u_{p}u_{2}+v_{p}v_{2})+
(u_{p}-v_{p})(u_{1}u_{2}+v_{1}v_{2})
-(u_{2}-v_{2})(u_{p}v_{1}+u_{1}v_{p}) \right)^{2}
\delta(E_{p}+E_{1}-E_{2}) \nonumber \\
&+&\left. \left( v_{p}u_{1}u_{2}+u_{p}u_{2}v_{1}+u_{1}v_{2}u_{p}-
v_{p}v_{2}u_{1}-v_{1}u_{2}v_{p}-u_{p}v_{2}v_{1} \right)^{2}
\delta(E_{p}+E_{1}+E_{2}) \right].
\label{eq:lowtc121}
\ea
The last term in (\ref{eq:lowtc121}) clearly vanishes because of the 
energy delta function. 
We recall that all {\it u}'s, {\it v}'s
and the quasiparticle energy $E_{p}$
in (\ref{eq:lowtc121}) have an implicit $({\bf R},T)$ dependence.  
If we change $\vp_{1}\rightarrow -\vp_{1}$ in the second term in 
(\ref{eq:lowtc121}), and use  (\ref{eq:minusf}), 
we can simplify (\ref{eq:lowtc121}) slightly to obtain
\ba
&&C_{12}[f]=2g^{2}n_{c}\co\int \frac{d\vp_{1}d\vp_{2}}{(2\pi)^{2}}
\left[(1+f)f_{1}f_{2}-f(1+f_{1})(1+f_{2})\right]\nonumber \\
&&\left[ \left( (u_{1}-v_{1})(u_{p}u_{2}+v_{p}v_{2})+
(u_{2}-v_{2})(u_{p}u_{1}+v_{p}v_{1})
-(u_{p}-v_{p})(u_{1}v_{2}+v_{1}u_{2}) \right)^{2}
\right.\nonumber \\
&&\times \delta(\vp-\vp_{1}-\vp_{2})\delta(E_{p}-E_{1}-E_{2})
\nonumber \\
&-&2\left( (u_{1}-v_{1})(u_{p}u_{2}+v_{p}v_{2})+
(u_{p}-v_{p})(u_{1}u_{2}+v_{1}v_{2})
-(u_{2}-v_{2})(u_{p}v_{1}+u_{1}v_{p}) \right)^{2}
\nonumber \\
&&\left. \times \delta(\vp+\vp_{1}-\vp_{2})
\delta(E_{p}+E_{1}-E_{2}) \right].
\label{eq:lowtc12}
\ea
The first term in (\ref{eq:lowtc12}) describes 
the decay of an excitation with momentum 
$\vp$ into two excitations with momenta
$\vp_{1}$ and $\vp_{2}$. At $T$ = 0, 
this is the only scattering process possible since there are no 
thermal excitations. The second term describes an 
excitation of momentum $\vp$ absorbing  
a thermal excitation of momentum  $\vp_{1}$, leaving an  
excitation with momentum $\vp_{2}=\vp+\vp_{1}$. 
This form of the collision integral was first 
written down by Eckern \cite{eckern}, and shortly after 
Kirkpatrick and  Dorfman \cite{kd} 
gave a more detailed derivation. 
Here, we have used the Kadanoff-Baym approach to
give a cleaner derivation of 
$C_{12}$, in a form which is also valid for 
a trapped Bose-condensed gas. 

After some algebra, one can also rewrite (\ref{eq:lowtc12}) in the 
following more compact form \cite{kd}
\ba
&&C_{12}[f]=2g^{2}n_{c}\co \int \frac{d\vp_{1}d\vp_{2}d\vp_{3}}
{(2\pi)^{2}}\mid A(2,3;1)\mid ^{2} \delta(\vp_{1}-\vp_{2}-\vp_{3})
\delta(E_{1}-E_{2}-E_{3}) \nonumber \\
&& \left[ \delta(\vp-\vp_{1})
-\delta(\vp-\vp_{2})-\delta(\vp-\vp_{3}) \right] 
\left[(1+f_{1})f_{2}f_{3}-f_{1}(1+f_{2})(1+f_{3})\right].
\label{eq:lowtc122}
\ea
Here the scattering amplitude in $|A|^{2}$ is
given in term of the Bose coherence 
factors {\it u} and {\it v} 
\be
A(2,3;1)\equiv (u_{3}-v_{3})(u_{1}u_{2}+v_{1}v_{2})+
(u_{2}-v_{2})(u_{1}u_{3}+v_{1}v_{3})
-(u_{1}-v_{1})(u_{2}v_{3}+v_{2}u_{3}) .
\label{eq:scattampli}
\ee
The first term in (\ref{eq:lowtc122}) is equivalent to the first term 
in (\ref {eq:lowtc12}), while the other term in (\ref{eq:lowtc12}) is 
equivalent to the second and third terms in (\ref{eq:lowtc122}).   

In conclusion,  the kinetic equation 
we have derived for thermally excited quasiparticles is given by
\ba
\left[\frac{\partial}{\partial T}+\nabla_{\vp}\left(E_{p}
+{\bf v}_{s}\cdot \vp\right)\cdot \nabla_{\bf R}
-\nabla_{\bf R}\left(E_{p}+{\bf v}_{s}\cdot\vp\right)
\cdot \nabla_{p}\right]f_{qp}(\vp,{\bf R},T) =C_{12}[f_{qp}] 
\label{eq:kinlowT}
\ea
with $C_{12}$ given explicitly by (\ref{eq:lowtc12}), or 
equivalently, (\ref{eq:lowtc122}).
The derivation of this equation is 
the main result of this paper. 

To remove the rapidly varying phase of the order parameter real, 
we have gauge transformed to the local rest frame
 where the condensate is at rest. Hence, the energy of
the thermally excited quasiparticles is measured relative to 
this local frame. Since the thermal excitations are moving  with the 
superfluid velocity ${\bf v}_{s}$ relative to the condensate, the 
energy of quasiparticles measured relative to the condensate is 
$E_{p}+{\bf v}_{s}\cdot \vp$. Therefore, the expression 
$E_{p}+{\bf v}_{s}\cdot \vp$ in the streaming term on the left-hand 
side of (\ref{eq:kinlowT}) 
is expected. Similarly, if we denote the quasiparticle 
distribution in the coordinate system where the quasiparticles are at 
rest by $f(\vp, \omega;{\bf R},T)$, the quasiparticles  
moving with the velocity  ${\bf v}_{s}$ relative to the 
condensate will be described by the distribution function 
$f(\vp, \omega-{\bf v}_{s}\cdot \vp   ; {\bf R},T)
\equiv f_{qp}(\vp,{\bf R},T)$, as occurs in  (\ref{eq:kinlowT})
 \cite{methodsqft,statmechlandau}. 

If we use the frame of reference where 
the quasiparticles are at rest, the streaming term will 
include the energy of the quasiparticles only ( i.e., the ${\bf 
v}_{s}\cdot\vp$ term in (\ref{eq:kinlowT}) will not be present). 
This lab frame of reference is used in the work 
of Zaremba, Nikuni and Griffin \cite{zngjltp}.

To understand the $C_{12}$ collision integral in 
(\ref{eq:lowtc12}) better and the 
corresponding scattering processes that it describes,
it is useful to consider 
a few limiting cases for a uniform gas.
We define $p_{0}^{2}\equiv 2m gn_{c}$ 
as the characteristic momentum 
for the crossover between the linear and the quadratic part of the 
quasiparticle spectrum ($p_{0}=\hbar k_{0}\equiv \hbar \xi^{-1}$, 
where $\xi$ is the healing 
length). We then consider the following special cases:

1) If all momenta $p_{i}\gg p_{0}$, then the quasiparticle spectrum 
$E_{p}$ defined by (\ref{eq:defquasie}) is equal to a single-particle 
spectrum $\tilde{\epsilon}_{p}$ given in (\ref{eq:defspe}). 
Moreover in this limit, 
 it follows from (\ref{eq:defuv}) that $u\rightarrow 1$ and 
$v\rightarrow 0$. Hence, the scattering amplitude $A$ in 
(\ref{eq:scattampli}) becomes unity 
and the collision integral in (\ref{eq:lowtc122}) then 
reduces to the one 
recently derived by Zaremba, Nikuni and Griffin \cite{itg2,zngjltp}
using a different approach. Clearly, 
this approximation is only valid at finite temperatures where 
the  dominant excitation spectrum is described by 
the Hartree-Fock 
single-particle spectrum in (\ref{eq:defspe}). 

2) In the opposite limit, when all three momenta $p_{i}$ 
are small, one can expand 
the Bose coherence factors {\it u} and {\it v} in 
the following way \cite{giorgini97,stringarilandau}
\ba
&&u_{p}\simeq \left(\frac{gn_{c}}{2E_{p}}\right)^{1/2}
+\frac{1}{2}\left(\frac{E_{p}}{2gn_{c}}\right)^{1/2}  
\nonumber \\
&&v_{p}\simeq \left(\frac{gn_{c}}{2E_{p}}\right)^{1/2}
-\frac{1}{2}\left(\frac{E_{p}}{2gn_{c}}\right)^{1/2}  
\label{eq:lowp}
\ea
where $E_{p}\simeq cp$ and $c=\sqrt{gn_{c}/m}$ is the speed of 
Bogoliubov sound. The sign of $v_{p}$ in 
(\ref{eq:lowp}) is opposite from the one given 
in \cite{giorgini97,stringarilandau} 
because we have defined $u_{p}v_{p}$ in (\ref{eq:defuv}) to be 
positive. In this limit, one obtains for the scattering 
amplitude the following expression \cite{eckern}
\be
A(1;2,3) \simeq \frac{3}{2^{7/4}}\sqrt{\frac
{p_{1}p_{2}p_{3}}{p_{0}^{3}}}
\ee 
This approximation is valid at low temperatures,  where only 
low-momentum excitations are relevant.  

3) Finally, one can consider the scattering of phonons (low-momentum 
excitations) with momentum  $p_{3} \ll p_{0}$ with particles 
(high-momentum excitations) with momenta $p_{1},p_{2}\gg p_{0}$. The 
corresponding amplitude for this process is given by \cite{eckern}
\be
A(1;2,3) \simeq 2^{3/4} (p_{3}/p_{0})^{1/2}.
\ee 
Therefore, in the case of a sound wave  scattering with particle-like 
excitations the scattering amplitude only depends on 
the wavevector of the sound wave. It 
is independent of the momenta of the scattering particles. 
  
\section{Local Equilibrium Solution}

To describe the thermalization of quasiparticles, 
it is sufficient to consider the $C_{22}$ collision integral.
From (\ref{eq:deff}), 
(\ref{eq:generalcoll}) and (\ref{eq:sigmabeliaevft}), one obtains
\ba
C_{22}[f]&=&-\frac{1}{2}g^{2}\int %\frac{d\omega}{2\pi}%
\frac{d{\bf p}_{i}d\omega_{i}}
{(2\pi)^{8}}\delta(E_{p}+\omega_{1}
-\omega_{2}-\omega_{3})\delta({\bf p}
+{\bf p}_{1}-{\bf p}_{2}-{\bf p}_{3}) 
\nonumber \\
&\times 
&\left[ff_{1}(1+f_{2})(1+f_{3})-(1+f)(1+f_{1})f_{2}f_{3}\right]
\nonumber \\
&&\left[ Tr\left(\hat{a}({\bf p}_{2},\omega_{2})
\hat{a}({\bf p},E_{p})\right)Tr\left(\hat{a}
({\bf p}_{1},\omega_{1})\hat{a}({\bf p}_{3},\omega_{3})\right)
\right. \nonumber \\
&+& \left. Tr\left(2\hat{a}({\bf p}_{2},\omega_{2})
\hat{a}({\bf p}_{1},\omega_{1})
\hat{a}({\bf p}_{3},\omega_{3})
\hat{a}({\bf p},E_{p}) \right)\right].
\label{eq:c22general}
\ea
The local equilibrium distribution function for  
quasiparticles $f_{0}(\vp,\omega;{\bf R},T)$ is determined by the 
requirement that $C_{22}[f_{0}]=0$. One can 
see from (\ref{eq:c22general}) that one doesn't have 
to specify some specific  approximation for 
the single-particle spectral densities.
 We only need a solution for $f$ such that 
the expression in (\ref{eq:c22general}) containing the 
$f$'s vanishes. One can verify that 
$C_{22}[f_{0}]=0$ if $f_{0}(\vp,\omega;{\bf R},T)$ 
has the following form
\be
f_{0}(\vp,\omega;{\bf R},T)=\frac{1}{e^{\beta\left(\omega-\vp 
\cdot ({\bf v}_{n}-{\bf v}_{s})-\mu_{qp}\co\right)}-1}.
\label{eq:eqquasi}
\ee
The vector ${\bf v}_{n}-{\bf v}_{s}$ describes the mean drift 
velocity of 
the quasiparticle gas in the local rest frame of the superfluid
and for small velocities we have \cite{kd,khalatnikov}
\be
\rho_{n}\left({\bf v}_{n}-{\bf v}_{s}\right)
=\int d{\vp} \vp f_{0}(\vp,\omega;{\bf R},T).
\label{eq:defrhonormal}
\ee
which defines the normal density $\rho_{n}$. 
This is consistent with the usual Landau definition. 
The distribution function given in (\ref{eq:eqquasi}) 
differs from the usual equilibrium quasiparticle 
distribution function discussed in the standard literature for 
phonons and rotons in liquid helium \cite{khalatnikov}. 
Since the number of quasiparticles 
in not conserved, the usual form for the equilibrium quasiparticle 
distribution 
function has no chemical potential (i.e., the chemical potential is 
zero). 
Here, we have introduced a chemical potential in  (\ref{eq:eqquasi}) 
to allow for the possibility that the condensate atoms and the 
thermally excited quasiparticles are {\it not} in the diffusive 
equilibrium with each other.  

To understand the physics of the quasiparticle chemical potential 
in (\ref{eq:eqquasi})
better, let us first  consider the high temperature case.
At high temperatures, the particle and the quasiparticle excitation 
spectrum are equivalent and the 
the local equilibrium distribution function in the lab frame is 
given by \cite{zngjltp}
\be
f_{0}(\vp,{\bf R},T)=\frac{1}{e^{\beta\left(\frac
{(\vp-m{\bf v}_{n})^{2}}{2m}+U\co-\tilde{\mu}\co\right)}-1}
\label{eq:flocalhighT}
\ee
with $U\co=U_{ext}({\bf R})+2gn\co$ and $\omega-{\bf v}_{s}\cdot 
\vp=E_{p}$ (see (\ref{eq:quasif})).
If we transform (\ref {eq:flocalhighT}) to the 
local rest frame (where ${\bf v}_{s}=0$), then $\vp'\equiv 
\vp-m{\bf v}_{s}$ is the momentum in the local rest frame 
and (\ref {eq:flocalhighT}) becomes
\be
f_{0}(\vp,{\bf R},T)=\frac{1}{e^{\beta\left[E_{p'}-\vp' 
\cdot ({\bf v}_{n}-{\bf v}_{s})-(\tilde{\mu}-\mu_{c})+
\frac{1}{2}m({\bf v}_{n}-{\bf v}_{s})^{2}\right]}-1}.
\ee  
Here $E_{p'}\equiv \frac{p'^{2}}{2m}+gn_{c}\co$ is the 
excitation energy  in the local rest frame. 
Therefore, if we define \cite{zngjltp} 
\be
\mu_{diff}\equiv \tilde{\mu}-\mu_{c}-
\frac{1}{2}m({\bf v}_{n}-{\bf v}_{s})^{2}, 
\ee
 we see the quasiparticle chemical potential $\mu_{qp}$ introduced in 
(\ref{eq:eqquasi}) can be identified with  $\mu_{diff}$ discussed at 
length by Zaremba, Nikuni and Griffin 
\cite{zngjltp}.
Hence, we see that in the quasiparticle description in a local frame 
in which the condensate is stationary, the difference ($\mu_{diff}$) 
between the chemical potentials of the condensate and non-condensate 
that was introduced by ZNG to describe the non-diffusive equilibrium 
of these two components appears very naturally 
as the quasiparticle chemical potential.  
The standard case discussed in 
the superfluid helium $^{4}$He 
literature \cite{khalatnikov} corresponds to $\mu_{qp}=0$ 
(see, however, the discussion of the second viscosity coefficients in 
superfluid $^{4}$He \cite{khalatnikov}). 

To summarize, we can distinguish two distribution functions $f$ which 
satisfy $C_{22}[f_{0}]=0$:

(1) The condensate atoms and the quasiparticle excitations are in  
diffusive thermal equilibrium, 
i.e. $\mu_{qp}=0$ and hence $C_{12}[f]=0$.

(2)The condensate atoms and the quasiparticle excitations are not in 
the 
diffusive thermal equilibrium, i.e. $\mu_{qp}\neq 0$. 
In this case, one finds that $C_{12}[f]$ is proportional to $
[1-e^{\beta \mu_{qp}}]$,
as in ZNG \cite{zngjltp}.

\section{Generalized Gross-Pitaevskii equation}

As we noted in the Introduction, the dynamics of 
a trapped Bose-condensed gas is usually described in the 
literature  by the Gross-Pitaevskii equation of motion 
(\ref{eq:gpeq}) for the condensate order parameter. 
Linearizing this equation, one obtains the collective 
mode frequencies that have been confirmed in many 
experiments \cite{strtheory}. 
However, at finite temperatures, the simple GP equation does not 
provide an adequate description of the thermally excited atoms.
Moreover, even at $T\ll T_{BEC}$, in recent 
experiments at JILA on ${}^{85}$Rb \cite{jilarb85},
the dimensionless gas parameter can be as large as 
$\sqrt{n_{c}a^{3}}\sim 
10^{-1}$, i.e. the quantum depletion of the condensate is not 
negligible. The simplest generalization of the zero temperature 
GP equation of motion is usually done by including  an 
additional self-consistent Hartree-Fock mean field 2g$\tilde{n}$ 
produced by the thermally excited atoms. The condensate atoms 
described by $\Phi\co$ move in this mean-field, in addition to the 
field produced by the condensate. However it is clear that the 
second-order collisions which we have included in deriving the 
quasiparticle kinetic equation in Section IV must also be included in 
a generalized GP equation for $\Phi\co$. Technically, this arises 
from  the three-field correlation function given in 
(\ref{eq:etadef}). 

Zaremba, Nikuni and Griffin \cite{zngjltp} have 
evaluated this three-field correlation function for the thermally 
excited atoms following the method of Kirkpatrick and 
Dorfman \cite{kd}, a method which  is not very transparent. 
ZNG obtained a generalized Gross-Pitaevskii 
equation with a dissipative term associated with
the $C_{12}$ collisions. This new dissipative 
term is, as expected, proportional 
to the collision integral for scattering between atoms in the 
condensate and thermal atoms, since such 
collisions change the number of atoms in the condensate.  
The ZNG work was limited to finite temperatures where the thermal 
atoms can be described as free atoms moving in the dynamic 
Hartree-Fock mean-field produced by all other atoms (both those in 
the 
condensate and in the thermal cloud). Recently, we have 
derived the same generalized GP equation as ZNG using 
the powerful KB method method. In this section,we now extend this 
kind of 
calculation to deal with low temperatures. The new 
equation of motion for the order parameter will be shown to be 
identical to that obtained in Refs. \cite{itg2,zngjltp} apart from 
the fact that $C_{12}$ is now given by the expression in 
(\ref{eq:lowtc12}).
That is, it now involves the Bogoliubov quasiparticles and 
collision cross-section is renormalized by various
 Bose-coherence factors involving the $u$'s and $v$'s.

To derive an equation of motion for the condensate order 
parameter, we first write equation (\ref{eq:explicitgp}) 
for $\Phi(\r,t)$ in the local rest frame. As before, under the gauge 
transformation (\ref{eq:gauge}), the only change is that the 
 non-interacting 
propagator is now given by (\ref{eq:g0llf}). The equation 
of motion in the new local frame is (see (\ref{eq:orderg1/2}))
\ba
\left[i\frac{\partial}{\partial t}
\right.&-&\left.\frac{\partial \theta(1)}
{\partial t}+\frac{1}{2m}\left[\nabla_{\r}
+im{\bf v}_{s}(1)\right]^{2}+
\mu_{0}-U_{ext}(\r)-g\left(2\tilde{n}(1)+n_{c}(1)
\right)\right]\Phi(1) \nonumber \\
&=&\int_{-\infty}^{t}d\bar{1}\left[S^{>}_{11}
-S^{<}_{11}\right]\left(\r-\bar{\r},t-\bar{t};
(\r+\bar{\r})/2,(t+\bar{t})/2\right)
\Phi(\bar{\r},\bar{t}) \nonumber \\
&+&\int_{-\infty}^{t}d\bar{1}\left[S^{>}_{12}
-S^{<}_{12}\right]\left(\r-\bar{\r},t-\bar{t};
(\r+\bar{\r})/2,(t+\bar{t})/2\right)
\Phi^{*}(\bar{\r},\bar{t}). \label{eq:orderpa}
\ea
Here, we have rewritten the condensate self-energy in the 
center-of-mass and relative coordinates and, as usual, set 
$\tilde{m}=0$ (the Popov approximation).
We recall that in the local rest frame, the 
order parameter phase is removed and hence 
$\Phi(\r,t)=\sqrt{n_{c}(\r,t)}$. 

We assume, as usual, that the $S$ correlation 
function (\ref{eq:sigmacondensate}) is 
dominated by small values of the relative space-time coordinates
$(\r-\bar{\r},t-\bar{t})$. Hence we can 
approximate $S^{\down}_{11}$ in (\ref{eq:orderpa}) by 
$S^{\down}(\r-\bar{\r},t-\bar{t};\r,t)$. For 
the same reason, we can also approximate the macroscopic 
wavefunction $\Phi(\bar{\r},\bar{t})$ in the integrand 
of (\ref{eq:orderpa}) by 
$\Phi(\r,t)\equiv \sqrt{n_{c}(\r,t)}$. 
Hence, (\ref{eq:orderpa}) simplifies to
\ba
&&\left[i\frac{\partial}{\partial t}
- \frac{\partial \theta(1)}
{\partial t}+\frac{1}{2m}\left[ \nabla_{\r}
+im{\bf v}_{s}(1)\right]^{2}+\mu_{0}-
U_{ext}(\r)-g\left(2\tilde{n}(1)+n_{c}(1)
\right)\right]\Phi(1) \nonumber \\
&=&\int_{-\infty}^{t}d\bar{\r}d\bar{t}
\left[\left(S^{>}_{11}-S^{<}_{11}\right)
\left(\r-\bar{\r},t-\bar{t};\r,t\right)
\Phi(\r,t)+\left(S^{>}_{12}-S^{<}_{12}\right)
\left(\r-\bar{\r},t-\bar{t};\r,t\right)\Phi^{*}(\r,t)
\right].
\label{eq:orderexact}
\ea
We can rewrite (\ref{eq:orderexact})  
(labeling $(\r,t)\rightarrow \co$) as follows
\ba
&&\left[i\frac{\partial}{\partial T}-
\frac{\partial \theta\co}{\partial T}
+\frac{1}{2m}\left[ \nabla_{\bf R}
+im{\bf v}_{s}\co\right]^{2}
+\mu_{0}-U_{ext}({\bf R}) \right. \nonumber \\
&-&\left.g\left(2\tilde{n}\co+n_{c}\co\right)
\right]\Phi\co\nonumber \\
&=&\Phi\co\int\frac{d{\bf p}d\omega}{(2\pi)^{4}}
\left[S^{>}_{11}-S^{<}_{11}+S^{>}_{12}-S^{<}_{12}\right]\pomrt
\int_{-\infty}^{T}d\bar{\r} d\bar{t} 
e^{i{\bf p}({\bf R}-\bar{\r})-i\omega (T-\bar{t})}.
 \label{eq:orderexact1}
\ea
In the second-order Beliaev approximation, the condensate 
self-energy is given by (\ref{eq:sigmacondensate}). 
The Fourier transform of this is 
\ba
\hat{S}^{\down}({\bf p},\omega;{\bf R},T)&
=&-\frac{1}{2}g^{2}\int \frac{d{\bf p}_{i}d\omega_{i}}
{(2\pi)^{8}}\delta(\omega+\omega_{1}
-\omega_{2}-\omega_{3})\delta({\bf p}
+{\bf p}_{1}-{\bf p}_{2}-{\bf p}_{3}) 
\nonumber \\
&\times &\left[ \tilde{g}^{\down}
({\bf p}_{2},\omega_{2};{\bf R},T)Tr\left[\tilde{g}^{\up}
({\bf p}_{1},\omega_{1};{\bf R},T)
\tilde{g}^{\down}
({\bf p}_{3},\omega_{3};{\bf R},T)\right]\right.
\nonumber \\ 
&+&\left. 2\tilde{g}^{\down}
({\bf p}_{2},\omega_{2};{\bf R},T)\tilde{g}^{\up}
({\bf p}_{1},\omega_{1};{\bf R},T)
\tilde{g}^{\down}
({\bf p}_{3},\omega_{3};{\bf R},T)\right].
\label{eq:sigmacondensateft}
\ea
In evaluating the right-hand side of 
(\ref{eq:orderexact1}), we use the identity
\be
\lim_{\delta \rightarrow 0^{+}}
\int_{-\infty}^{T}d\bar{t} 
e^{-i(\omega+i\delta )(T-\bar{t})}\simeq \pi 
\delta(\omega)+iP\left(\frac{1}{\omega}\right),
\ee 
and only keep the delta function part, to obtain
\ba
&&\left[i\frac{\partial}{\partial T}-
\frac{\partial \theta\co}{\partial T}
+\frac{1}{2m}\left[ \nabla_{\bf R}
+im{\bf v}_{s}\co\right]^{2}
+\mu_{0}-U_{ext}({\bf R})-g\left(2\tilde{n}\co
+n_{c}\co\right) \right. \nonumber \\
&-&\left.\left(S^{>}_{11}-S^{<}_{11}+S^{>}_{12}-S^{<}_{12}\right)
(\vp=0,\omega=0;{\bf R},T)\right]\Phi\co=0.
\label{eq:orderexact2}
\ea

Using (\ref{eq:deff}) in the  condensate self-energy $S$ 
given in (\ref{eq:sigmacondensateft}),
the second-order terms appearing in (\ref{eq:orderexact2}) 
reduce to 
\ba
\left(S^{>}_{11}-S^{<}_{11}\right.&+&\left.
S^{>}_{12}-S^{<}_{12}\right)
(\vp=0,\omega=0;{\bf R},T)= 
i\frac{1}{2}g^{2}\int \frac{d{\bf p}_{i}d\omega_{i}}
{(2\pi)^{8}}\delta(\omega_{1}
-\omega_{2}-\omega_{3})\delta(
{\bf p}_{1}-{\bf p}_{2}-{\bf p}_{3}) 
\nonumber \\
&\times &
\left[f_{1}(1+f_{2})(1+f_{3})-(1+f_{1})f_{2}f_{3}\right]
\left[ (a_{11}+a_{12})({\bf p}_{2},\omega_{2})
Tr\left(\hat{a}({\bf p}_{1},\omega_{1})
\hat{a}({\bf p}_{3},\omega_{3})\right)\right.
\nonumber \\ 
&+&\left. 2\left(\hat{a}({\bf p}_{2},\omega_{2})
\hat{a}({\bf p}_{1},\omega_{1})
\hat{a}({\bf p}_{3},\omega_{3})\right)_{11}+
2\left(\hat{a}({\bf p}_{2},\omega_{2})
\hat{a}({\bf p}_{1},\omega_{1})
\hat{a}({\bf p}_{3},\omega_{3})\right)_{12}
\right].
\ea
Recalling that in the local rest frame we have  
$\Phi \co=\sqrt{n_{c}\co}$, with no phase,
we finally obtain a generalized 
Gross-Pitaevskii equation in the following form
\ba
&&i\frac{\partial \sqrt{n_{c}\co}}{\partial T}=
\left[\frac{\partial \theta\co}{\partial T}
-\frac{1}{2m}\left[ \nabla_{\bf R}+im{\bf 
v}_{s}\co\right]^{2}-\mu_{0}\right. \nonumber \\
&+&\left.U_{ext}({\bf R})+g\left[
2\tilde{n}\co+n_{c}\co\right]-iR\co
\right]\sqrt{n_{c}\co}. 
\label{eq:gengp}
\ea
The new dissipative term {\it R} in the GP 
equation is clearly related to the $C_{12}$ collision 
term in the kinetic equation (\ref{eq:kinlowT}), 
namely \cite{zngjltp,itg2}
\be
R\co\equiv \int \frac{d{\bf p}}{(2\pi)^{3}}
\frac{C_{12}[f\cof]}{2n_{c}\co}.
\ee
This term describes the damping of condensate 
amplitude fluctuations due to collisions with 
the thermal excitations. The appearance 
of the dissipative term in (\ref{eq:gengp}) is 
expected since the $C_{12}$ collisions change 
the number of atoms in the condensate and hence 
can modify the magnitude of the condensate macroscopic 
wavefunction. We note that since we ignore the  
real part of the second-order 
self-energies, the condensate chemical potential in 
(\ref{eq:chemicalpot}) is not modified.  
If we transform back into the lab frame (where we have 
$\Phi=\sqrt{n_{c}}e^{i\theta}$), (\ref{eq:gengp}) 
reduces to the time-dependent generalized 
Gross-Pitaevskii equation for $\Phi\co$
discussed by ZNG \cite{zngjltp}. However,
$C_{12}$ now involves the Bogoliubov 
quasiparticle spectrum  in 
place of the HF particle-like spectrum used in Ref. \cite{zngjltp}, 
and in addition, the collision integral matrix elements involve the 
characteristic Bose coherence factors $u$ and $v$.

\section{Kohn mode}
\newcommand{\ta}{\mbox{\boldmath $\eta$}}
In this section, we show that the non-condensate and condensate 
both exhibit the rigid in-phase oscillations, the Kohn mode.
This mode is discussed in detail in Section VI of \cite{zngjltp}, and 
the 
analysis there is easily generalized to the more general equations we 
are 
discussing. The center-of-mass oscillation of 
the non-condensate and condensate 
density profiles corresponding to the Kohn mode is given by
\ba
n_{c}\co&\equiv& n_{c0}({\bf R}-\ta(T)) \nonumber \\
\tilde{n}\co&\equiv& \tilde{n}_{0}({\bf R}-\ta(T)).
\label{eq:kohndensity}
\ea
Here, the center-of-mass displacement $\ta(T)$ 
(with ${\bf v}_{s}=\dot {\ta})$ satisfies the 
harmonic oscillator equation of motion
\be
m\frac{\partial^{2}\eta_{\alpha}}{\partial T^{2}}=
-\omega_{\alpha}^{2}\eta_{\alpha},
\ee
where $\omega_{\alpha}$ is the trap frequency in the 
$\alpha$${}^{th}$ 
direction. The quasiparticle distribution function $f(\vp,{\bf 
R},T)$ corresponds to the equilibrium density profile oscillating 
around its center of mass with the trap frequency, i.e., 
\be
f(\vp,{\bf R},T)\equiv f_{0}(\vp,{\bf R}-\ta(T)).
\label{eq:kohneqdistr}
\ee

To prove (\ref{eq:kohneqdistr}), we note that with 
(\ref{eq:kohndensity}), 
the expression for the Bogoliubov excitation energy in 
(\ref{eq:defquasie}) reduces to 
\be
E_{p}\co= \sqrt{\epsilon_{p}^{2}+2gn_{c0}({\bf R}
-\ta)\epsilon_{p}}\equiv E_{p0}({\bf R}-\ta).
\ee 
Therefore, the kinetic equation for the 
quasiparticle distribution function in (\ref{eq:kinlowT}) is
\ba
\left[\frac{\partial}{\partial 
T}\right.&+&\left.\nabla_{\vp}\left(E_{p0}
({\bf R}-\ta(T))+\dot{\ta} \cdot \vp\right)\cdot 
\nabla_{\bf R}
-\nabla_{\bf R} E_{p0}({\bf R}-\ta(T))
\cdot \nabla_{p}\right]f_{0}(\vp,{\bf R}-\ta (T)) \nonumber \\
&=&C_{12}[f_{0}(\vp,{\bf R}-\ta(T))]. 
\label{eq:kohnkinetic1}
\ea
If we expand $f_{0}(\vp,{\bf R}-\ta (T))$ around 
$\ta =0$,
\be
f_{0}(\vp,{\bf R}-\ta(T))=f_{0}(\vp,{\bf R})-\ta\cdot
\nabla_{\bf R}f_{0}(\vp,{\bf R}),
\ee
and neglect the quadratic terms in $\eta$, 
(\ref{eq:kohnkinetic1}) simplifies to 
\ba
\nabla_{\vp}E_{p0}({\bf R}-\ta (T))\cdot \nabla_{\bf R}
f_{0}(\vp,{\bf R}-\ta(T))
&-&\nabla_{\bf R} E_{p0}({\bf R}-\ta (T))
\cdot \nabla_{p}f_{0}(\vp,{\bf R}-\ta (T))
\nonumber \\
&=&C_{12}[f_{0}(\vp,{\bf R}-\ta(T))]. 
\label{eq:kohnkinetic2}
\ea
The left-hand side of (\ref{eq:kohnkinetic2}) is seen to be 
the kinetic equation 
for the equilibrium distribution function. To prove that the 
Kohn mode is a solution, one only has to show 
$C_{12}[f_{0}(\vp,{\bf R}-\ta(T)]=0$. Assuming the 
equilibrium quasiparticle distribution function $f_{0}$ 
is given by (\ref{eq:eqquasi}) with $\mu_{qp}=0$, 
and using the identity for 
the Bose distribution function
\be
1+f(x)=e^{x}f(x),
\ee
one obtains following expression for the $C_{12}[f_{0}]$
\ba
&&C_{12}[f_{0}(\vp,{\bf R}-\ta(T))]
=2g^{2}n_{c}({\bf R}-\ta(T)) \int 
\frac{d\vp_{1}d\vp_{2}d\vp_{3}}
{(2\pi)^{2}}\mid A_{0}(2,3;1)\mid ^{2} 
\nonumber \\
&&\times \delta(\vp_{1}-\vp_{2}-\vp_{3})
\delta\left[E_{10}({\bf R}-\ta(T))-E_{20}({\bf R}-\ta(T))
-E_{30}({\bf R}-\ta (T))\right] \nonumber \\
&&\times  \left[ \delta(\vp-\vp_{1})
-\delta(\vp-\vp_{2})-\delta(\vp-\vp_{3}) \right] 
(1+f_{10})f_{20}f_{30} \nonumber \\
&&\times \left[1-e^{-\beta\left[E_{10}-E_{20}-E_{30}
-(\vp_{1}-\vp_{2}-\vp_{3})({\bf v}_{n}-{\bf v}_{s})\right]}\right].
\label{eq:kohnc12}
\ea
Using the delta functions in (\ref{eq:kohnc12}) corresponding to 
the conservation of energy and momentum, it immediately follows 
that     
\be
C_{12}[f_{0}(\vp,{\bf R}-\ta (T))]=0.
\ee
This proves that the non-condensate exhibits a rigid simple harmonic 
displacement with the trap frequency.       

Since we have proven that the collision integral vanishes for the 
Kohn mode type of oscillations, the dissipative term $R$ in the 
generalized GP equation in (\ref{eq:gengp}) 
will vanish as well. Therefore, it follows that 
the equilibrium condensate profiles oscillate with the trap frequency 
(for the analogous calculation based on the HF and HFB 
single-particle 
spectrum, see Refs. \cite{zngjltp,itg1}).

\section{Conclusions}
Using the Kadanoff-Baym non-equilibrium Green's 
function formalism \cite{kb,kk} we have derived a kinetic equation 
for 
the quasiparticle distribution function and a generalized 
Gross-Pitaevskii equation valid at all temperatures. 
Our new kinetic equation involves the Bogoliubov quasiparticle 
spectrum and Bose coherence factors involving the {\it u} and {\it v}
functions. Our equations reduce to those obtained at high 
temperatures by ZNG 
\cite{zngjltp}. As we have emphasized in Section II, the 
approximation 
that we have used in this paper is gapless and it gives the correct 
low-momentum (long-wavelength) limit.
In contrast, the so-called ``conserving approximations''  
are based on a functional from which both self-energy 
$\Sigma$ and the source $\eta$ functions  can be 
derived by functional differentiation (see p. 338 ff of Ref. 
\cite{hm}, and also Section III in Ref. \cite{itg2}) . The resulting 
single-particle Green's function can be used to 
generate a density response function whose spectrum 
is guaranteed to satisfy conservation 
laws \cite{hfb,hm,kbphyrev,baym}, even though 
the generating Green's functions have an energy gap in the 
long-wavelength limit.
 
The simple quasiparticle approximation that we have used in this 
paper has allowed us to derive the kinetic 
equation in a ``Boltzmann-like form''.   
The kinetic equation in the Bogoliubov-Popov approximation given in 
(\ref{eq:kinlowT}) is only valid in this quasiparticle approximation. 
In deriving these results, 
we have neglected the real part of the 
second-order self-energies that give rise to many-body corrections.
Our discussion could be 
generalized to include these real parts, but this 
improved theory would be very complex. A first step would be to 
include such renormalization effects within a simple quasiparticle 
approximation to the spectral densities $a_{\alpha \beta}$.

Our kinetic equation for the quasiparticle distribution 
and the generalized 
Gross-Pitaevskii equation are coupled and have to be solved 
self-consistently. They should provide a sound basis 
for the future systematic study of the 
non-equilibrium response of a trapped Bose gas at low temperatures.
We remark that we could use our results to derive the 
Landau-Khalatnikov two-fluid hydrodynamic equations, in the 
collision-dominated region. Indeed, the 
approach developed in  Ref. \cite{khalatnikov} is based on a 
quasiparticle kinetic equation which is precisely of the kind we have 
derived in (\ref{eq:kinlowT}). 
Such a calculation would extend a recent derivation \cite{nikuninew} 
of the 
Landau-Khalatnikov two-fluid equations in the high temperature 
region where the simple HF single-particle spectrum is appropriate. 

%\section*{ACKNOWLEDGMENTS}
\begin{acknowledgements}
We would like to thank T. Nikuni, J. Williams and E. Zaremba for 
useful discussions related to this research. In particular, 
discussions with E. Zaremba clarified the relation between the  
equilibrium distribution function in the lab and local rest frame, as 
discussed in Section V. 
We also thank A.E. Jacobs for providing 
us with a copy of Kane's Ph.D. thesis (Ref. \cite{kt}). 
This research was supported by NSERC of Canada.
\end{acknowledgements}
	
\begin{appendix}
\section{}
For illustration, we give the detailed proof that the collision 
integral $I$ in (\ref{eq:generalcoll}) conserves momentum, namely
\be
\int d\vp \vp I[f(\vp,{\bf R},T)]=0.
\label{eq:appendixproof}
\ee
The proof is essentially the same as one uses in classical gases (See 
Ch. 5 of \cite{huang}).
Using the expression for the second-order Beliaev energy in 
(\ref{eq:sigmabeliaevft}) in conjunction with (\ref{eq:deff}), 
one obtains 
\ba
&&\int d\vp \vp I[f(\vp,{\bf R},T)]
=-\frac{1}{2}g^{2}\int \frac{d\vp d\omega}{(2\pi)^{4}}
\frac{d{\bf p}_{i}d\omega_{i}}
{(2\pi)^{8}}\delta(\omega+\omega_{1}
-\omega_{2}-\omega_{3})\delta({\bf p}
+{\bf p}_{1}-{\bf p}_{2}-{\bf p}_{3}) 
\nonumber \\
&&\vp \left[\left[ff_{1}(1+f_{2})-(1+f)(1+f_{1})f_{2}f_{3}\right]
\left[Tr\left(\hat{a}({\bf p}_{2},\omega_{2})
\hat{a}({\bf p},\omega)\right)Tr\left(\hat{a}({\bf p}_{1},\omega_{1})
\hat{a}({\bf p}_{3},\omega_{3})\right)+ \right. \right.\nonumber \\
&&\left. 2Tr\left(\hat{a}({\bf p}_{2},\omega_{2})
\hat{a}({\bf p}_{1},\omega_{1})\hat{a}
({\bf p}_{3},\omega_{3})\hat{a}({\bf p},\omega)
\right)\right] +
\label{eq:appendix1}\\
&&\left[f(1+f_{2})(1+f_{3})-(1+f)f_{2}f_{3}\right]
\left[ Tr\left(\hat{a}({\bf p}_{2},\omega_{2})
\hat{a}({\bf p},\omega)\right)
Tr\left(\hat{h}({\bf p}_{1},\omega_{1})
\hat{a}({\bf p}_{3},\omega_{3})\right)
+\right. \nonumber \\
&&\left. 2Tr\left(\hat{a}({\bf p}_{2},\omega_{2})
\hat{h}({\bf p}_{1},\omega_{1})\hat{a}
({\bf p}_{3},\omega_{3})\hat{a}({\bf p},
\omega)\hat{a}({\bf p},\omega)\right)\right]+
\label{eq:appendix2}\\
&&\left[ff_{1}(1+f_{2})-(1+f)(1+f_{1})f_{2}\right]
\left[ Tr\left(\hat{a}({\bf p}_{2},\omega_{2})
\hat{a}({\bf p},\omega)\right)
Tr\left(\hat{a}({\bf p}_{1},\omega_{1})
\hat{h}({\bf p}_{3},\omega_{3})\right)+
\right. \nonumber \\
&&\left. 2Tr\left(\hat{a}({\bf p}_{2},\omega_{2})
\hat{a}({\bf p}_{1},\omega_{1})\hat{h}
({\bf p}_{3},\omega_{3})\hat{a}({\bf p},
\omega)\hat{a}({\bf p},\omega)\right)\right]+
\label{eq:appendix3}\\
&&\left[ff_{1}(1+f_{3})-(1+f)(1+f_{1})f_{3} \right]
\left[ Tr \left( \hat{h}({\bf p}_{2},\omega_{2})
\hat{a}({\bf p},\omega)\right)
Tr\left(\hat{a}({\bf p}_{1},\omega_{1})
\hat{a}({\bf p}_{3},\omega_{3})\right)+ \right.
\nonumber \\
&&\left. \left. 2Tr \left(
\hat{h}({\bf p}_{2},\omega_{2})
\hat{a}({\bf p}_{1},\omega_{1})
\hat{a}({\bf p}_{3},\omega_{3})
\hat{a}({\bf p},\omega)\hat{a}({\bf p},\omega) \right) 
\right]\right]
\label{eq:appendix4}
\ea
Consider the (\ref{eq:appendix1}) term first. The change of dummy 
variables
\be
(\vp,\omega) \stackrel{\leftarrow}{\rightarrow} (\vp_{2},\omega_{2})
\hspace{5mm}(\vp_{1},\omega_{1}) 
\stackrel{\leftarrow}{\rightarrow} (\vp_{3},\omega_{3})
\label{eq:change1}
\ee
doesn't change the delta functions. Because of the cyclic invariance 
of 
trace, the first term in (\ref{eq:appendix1}) is also unchanged - the 
only change is that $\vp$ in front of the integral becomes 
$-\vp_{2}$. After the change of variables given by (\ref{eq:change1}),
the second term in (\ref{eq:appendix1}) becomes
\be
Tr\left(\hat{a}({\bf p},\omega)
\hat{a}({\bf p}_{3},\omega_{3})
\hat{a}({\bf p}_{1},\omega_{1})
\hat{a}({\bf p}_{2},\omega_{2})\right)=
Tr\left(\hat{a}({\bf p}_{2},\omega_{2})
\hat{a}({\bf p},\omega)
\hat{a}({\bf p}_{3},\omega_{3})
\hat{a}({\bf p}_{1},\omega_{1})\right)
\label{eq:appendix5}
\ee
Transforming $(\vp,\omega) \stackrel{\leftarrow}{\rightarrow} 
(\vp_{1},\omega_{1})$ in (\ref{eq:appendix5}), doesn't 
change either the $\delta$-functions or the  product of $f$'s. 
Therefore the trace in second term 
in (\ref{eq:appendix1}) doesn't change. 
Thus (\ref{eq:appendix1}) after the transformations is unchanged, 
but $\vp$ is replaced with $-\vp_{2}$. 
Using the transformation,
\be
(\vp,\omega) \stackrel{\leftarrow}{\rightarrow} (\vp_{1},\omega_{1})
\hspace{5mm} (\vp_{2}¥,\omega_{2}) 
\stackrel{\leftarrow}{\rightarrow} (\vp_{3},\omega_{3})
\label{eq:change2}
\ee
we would again obtain the same expression as (\ref{eq:appendix1}), 
but 
with $\vp$ replaced with $\vp_{1}$. Finally, if we make the 
transformation
\be
(\vp,\omega) \stackrel{\leftarrow}{\rightarrow} (\vp_{3},\omega_{3})
\hspace{5mm} (\vp_{2}¥,\omega_{2}) 
\stackrel{\leftarrow }{\rightarrow} (\vp_{1},\omega_{1})
\label{eq:change3}
\ee
we obtain the same expression as in (\ref{eq:appendix1}), but 
with $\vp$ replaced with $-\vp_{3}$. We conclude that  we can write 
(\ref{eq:appendix1}) as one fourth of the sum of four equivalent 
terms. The integrand of this new expression is thus seen to be 
proportional to
\be
\left(\vp+\vp_{1}-\vp_{2}-\vp_{3}\right)
\delta\left(\vp+\vp_{1}-\vp_{2}-\vp_{3}\right),
\ee
which clearly vanishes. 

A similar discussion can be given of the other terms in 
(\ref{eq:appendix2})
-(\ref{eq:appendix4}). Making the change of variables
\be
(\vp,\omega) \stackrel{\leftarrow}{\rightarrow} (\vp_{1},\omega_{1})
\hspace{5mm} (\vp_{2}¥,\omega_{2}) 
\stackrel{\leftarrow}{\rightarrow} (\vp_{3},\omega_{3})
\label{eq:change4}
\ee
in (\ref{eq:appendix3}) first, and 
using the cyclic invariance of the trace, one can show that 
(\ref{eq:appendix3}) is the same as (\ref{eq:appendix4}), but with 
$\vp_{1}$ instead of $\vp$ as an overall multiplying factor.
Using
\be
(\vp,\omega) \stackrel{\leftarrow}{\rightarrow} (\vp_{3},\omega_{3})
\hspace{5mm} (\vp_{2},\omega_{2}) 
\stackrel{\leftarrow}{\rightarrow} (\vp_{2},\omega_{2}).
\label{eq:change5}
\ee
in (\ref{eq:appendix2}), and then relabeling $\vp \rightarrow 
\vp_{1}$ in the second term in (\ref{eq:appendix2}),
we can reduce (\ref{eq:appendix2}) to the same expression as 
in (\ref{eq:appendix4}) but with $\vp$ replaced with $-\vp_{3}$. 
Therefore the sum of the terms 
(\ref{eq:appendix2}), (\ref{eq:appendix3}) and (\ref{eq:appendix4}) 
has an integrand involving
\be
\left(\vp+\vp_{1}-\vp_{3}\right)\delta\left(
\vp+\vp_{1}-\vp_{2}-\vp_{3}\right)
\delta(\vp_{2})\delta(\omega_{2}),
\ee
which clearly vanishes. This completes the proof of 
(\ref{eq:appendixproof}) and hence of momentum conservation by 
collisions.

\end{appendix}

\end{document}